\documentclass[a4paper,11pt]{article}
\pdfoutput=1 
\usepackage{jcappub} 
\usepackage[T1]{fontenc} 
\usepackage{array}
\usepackage{booktabs}
\usepackage{multirow}
\usepackage[normalem]{ulem}
\usepackage{placeins}
\usepackage[usenames,dvipsnames]{xcolor}

\usepackage{soul}

\DeclareMathOperator\erf{erf}

\newcommand{\beq}{\begin{equation}}
\newcommand{\eeq}{\end{equation}}
\newcommand{\bal}{\begin{align}}
\newcommand{\eal}{\end{align}}
\newcommand{\ns}{{N_\text{side}}}
\newcommand{\np}{{N_\text{pix}}}
\newcommand{\bs}[1]{\boldsymbol{#1}}

\usepackage{orcidlink}

\title{Fréchet Vectors as sensitive tools for blind tests of CMB anomalies}

\author[a]{Ricardo G. Rodrigues\orcidlink{0000-0003-3824-5524}}
\author[a, b, e]{Thiago S. Pereira\orcidlink{0000-0002-6479-364X}}
\author[b, c, d, e]{Miguel Quartin\orcidlink{0000-0001-5853-6164}}

\affiliation[a]{Departamento de Física, Universidade Estadual de Londrina, Rod. Celso Garcia Cid, Km
380, 86057-970, Londrina, Paraná, Brazil}
\affiliation[b]{Instituto de Física, Universidade Federal do Rio de Janeiro, 21941-972, Rio de Janeiro, RJ, Brazil}

\affiliation[c]{Observatório do Valongo, Universidade Federal do Rio de Janeiro, 20080-090, Rio de Janeiro, RJ, Brazil}

\affiliation[d]{PPGCosmo, Universidade Federal do Espírito Santo, 29075-910, Vitória, ES, Brazil}

\affiliation[e]{Centro Brasileiro de Pesquisas Físicas, 22290-180, Rio de Janeiro, RJ, Brazil}

\emailAdd{ricardo.gonzatto11@uel.br}
\emailAdd{tspereira@uel.br}
\emailAdd{mquartin@if.ufrj.br}

\abstract{Cosmological data collected on a sphere, such as CMB anisotropies, are typically represented by the spherical harmonic coefficients, denoted as $a_{\ell m}$. The angular power spectrum, or $C_\ell$, serves as the fundamental estimator of the variance in this data. Alternatively, spherical data and their variance can also be characterized using Multipole Vectors (MVs) and the Fréchet variance. The vectors that minimize this variance, known as Fréchet Vectors (FVs), define the center of mass of points on a compact space, and are excellent indicators of statistical correlations between different multipoles. We demonstrate this using both simulations and real data. Through simulations, we show that FVs enable a blind detection and reconstruction of the location associated with a mock Cold Spot anomaly introduced in an otherwise isotropic sky. Applying these tools to the 2018 Planck maps, we implement several improvements on previous null tests of Gaussianity and statistical isotropy, down to arc-minute scales. Planck's MVs appear consistent with these hypotheses  at scales $2 \leq\ell \leq 1500$ when the common mask is applied, whereas the same test using the FVs rejects them with significances between 5.3 and 8.2$\sigma$. The inclusion of anisotropic noise simulations render the FVs marginally consistent ($\geq 2\sigma$) with the null hypotheses at the same scales, but still rejects them at $3.5-3.7\sigma$ when we consider scales above $\ell=1500$, where the signal-to-noise is small. Limitations of the noise and/or foregrounds modeling  may account for these deviations from the null hypothesis.}

\begin{document}
\maketitle
\flushbottom

\section{Introduction}\label{sec:intro}

Ever since its discovery, cosmic microwave background (CMB) radiation has played a central role in the development of modern cosmology. Following its first detection by Penzias \& Wilson, the three generations of space-based instruments, together with several ground-based telescopes and balloon experiments (see~\cite{Bucher:2015eia} for a brief review), have cemented our understanding of the early universe and the Big Bang paradigm. In particular, they were pivotal in establishing $\Lambda$CDM as our \textit{de-facto} standard cosmological model.

CMB anisotropies are arguably the cleanest and most direct observational window into the early universe that we have. This conclusion is supported by our precise understanding of the physics that produce the observed fluctuations across different scales, due in part to the fact that these fluctuations largely occur in a regime where the dynamics is essentially linear. If the universe is further assumed to have primordial fluctuations which are Gaussian, and to have translational and rotational symmetries, as dictated by the Cosmological Principle, the angular power spectrum, $C_\ell$, becomes a summary statistic that encapsulates all available information. Although the measured $C_\ell$s of the CMB temperature map are in excellent agreement with what one would expect from a Gaussian and statistically isotropic (GSI) universe~\cite{WMAP:2003elm,Planck:2018vyg}, and no detection of primordial non-Gaussianity has yet been made~\cite{Planck:2019kim}, deviations from the GSI expectations are certainly present in the data. Some arise naturally from known second-order perturbations, such as CMB lensing~\cite{Hu:2000ee}, detected at over 40$\sigma$~\cite{Planck:2018lbu}, and aberration and Doppler couplings due to the observer motion~\cite{Challinor:2002zh,Notari:2011sb}, detected at around $6\sigma$~\cite{Ferreira:2020aqa}. However, other so-called anomalies have been found, which have no consensual explanation~\cite{Schwarz:2015cma,Planck:2015igc}. Some of these deviations may be due to different cosmological models, or they may be simpler flukes in the data.  Thus, a central issue is to tell whether measured deviations are of a cosmological, astrophysical, or systematic nature. This prompts us to examine the data using alternative mathematical representations, summary statistics, or both.

This particular approach to the analysis of CMB anisotropies is not new. In fact, CMB studies have long benefited from alternative representations of the usual harmonic analysis, as exemplified by the extensive use of wavelets~\cite{Hobson:1998av,Tenorio:1999mf,Cayon:2001cq,Vielva:2001sc,Aghanim:2003fs,Mukherjee:2004in}, 
Minkowski functionals~\cite{Schmalzing:1997uc,Eriksen:2004df,Hikage:2006fe,Novaes:2016wyx}, and multipole vectors~\cite{Copi:2003kt,Helling:2006xh,Bielewicz:2008ga,Pinkwart:2018nkc,Oliveira:2018sef}. These are versatile tools, applicable to a wide range of topics in CMB data analysis, from component separation methods~\cite{Planck:2018yye}, to non-Gaussianities~\cite{Eriksen:2004df,Hikage:2006fe,Aghanim:2003fs}, and cosmic topologies~\cite{Aurich:2006vt,Bielewicz:2008ga}. Alternatively, if a null test of the GSI hypotheses is desired, then one can keep the usual harmonic approach and dispose of anisotropic and model-independent implementations of two-point correlation functions and their estimators~\cite{Hajian:2004zn,Hajian:2005jh,Pullen:2007tu,Pereira:2009kg,Abramo:2009fe,Froes:2015hva}.

Among these tools, Multipole Vectors (MVs) figure as the least explored by cosmologists. Originally introduced by Maxwell in the context of electrostatics, they are an alternative to the standard harmonic basis, summarized by the multipolar coefficients $a_{\ell m}$s, to describe CMB fluctuations~\cite{Copi:2003kt}. In the multipole vector basis, the $2\ell+1$ degrees of freedom characterizing a given CMB multipole $\ell$ are split into $\ell$ unit headless vectors plus a scalar which, in the case of GSI skies, is proportional to the (cosmology-dependent) $C_\ell$s. Thus, in the standard cosmological model, MVs carry all the $C_\ell$-independent data contained in the $a_{\ell m}$s. Moreover, they are naturally invariant under spatial rotations, unlike the $a_{\ell m}$s which mix $m$s in such cases. These properties make the MVs an
interesting tool to probe deviations from Gaussianity and isotropy hypotheses in a model-independent way~\cite{Oliveira:2018sef}. This motivation led to a recent development of the code \texttt{polyMV}, discussed below, which allows efficient computation of the MVs at all scales, which was previously not possible~\cite{Oliveira:2018sef}.

Although the MV set contains all the information of a given scalar map of the sky, often in cosmology we make use of summary statistics to analyze the data. In the context of MVs, summary statistics have been designed for the specific task of detecting CMB anomalies~\cite{Copi:2003kt,Abramo:2006gw} or, more specifically, possible alignments between low $\ell$ multipoles, which is one of the observed anomalies of the CMB~\cite{Tegmark:2003ve,Schwarz:2004gk,Notari:2015kla}. For this reason, these statistics invariably mix vectors from different multipoles, and thus cannot be used as a summary of the behavior of the CMB at each angular scale. It is therefore interesting to have a summary statistic for each multipole, that is still capable of detecting deviations from isotropy. One such statistic is the Fréchet vectors (FVs), originally proposed in~\cite{Oliveira:2018sef}.

Frechét vectors borrow from the mathematical definition of variance in metric spaces, also known as Fréchet variance, to build a geometrically motivated and model-independent summary statistic of the MVs which results in one \textit{unit and headless vector} per CMB multipole. These vectors are geometrically defined as the position of the ``center of mass'' of the MVs on the unit sphere and have many interesting geometrical and statistical properties. First, being based on the MVs, they are $C_\ell$-independent and have an isotropic 1-point distribution function in the case of GSI skies~\cite{Oliveira:2018sef}. Second, since they result from compressing $\ell$ vectors into just one, they are less prone to cosmic variance. Finally, because the direction of each FV is defined by the collection of $\ell$ multipole vectors, the FVs are more susceptible to very small angular variations of each MV, which can result in greater sensitivity to anisotropies in the data. 

Here we demonstrate, using simulations, that the FVs can be directly correlated with spatial anisotropies artificially introduced in the input maps. This allows in principle for a blind reconstruction of spatial anisotropies of real CMB maps. We illustrate this for the case of a mock Cold Spot anomaly~\cite{Vielva:2003et,Owusu:2022etl}, and show that in the case of a spot of nearly the same aperture but twice as cold, FVs can detect its presence and pinpoint its axis blindly. 

We also analyze real Planck 2018 data with a straightforward chi-squared test of the null GSI hypotheses using 1-point statistics of both MVs and FVs. Overall, we find Planck's MVs to be consistent with GSI simulations, including mask and instrumental anisotropic noise across all scales considered ($2\leq\ell\leq2000$). In contrast, Planck's FVs rule them out with varying statistical strengths. These depend on whether we probe scales with high ($\ell\leq1500$) or low ($\ell>1500$) signal-to-noise ratios. Conservatively, we find \texttt{NILC} and \texttt{SMICA} to be marginally inconsistent with GSI simulations at $\gtrsim2\sigma$ and $\gtrsim3\sigma$, respectively, in the former region. The higher sensitivity of the FVs in comparison to the MVs is due to the construction of our chi-square test, and not an intrinsic properties of these vectors. This is explained in detail in Appendix~\ref{app:sens_frechet}.

We include many important improvements in the analysis compared to~\cite{Oliveira:2018sef}, such as a pixel-based implementation (as opposed to an independent analysis on both angular coordinates of these vectors), an improved evaluation of MVs and FVs covariance matrix, and the inclusion of the instrumental noise in our tests, using Planck's anisotropic noise simulations, which allowed for an extension of our tests to the scales $1500\leq\ell\leq2000$, where CMB maps are known to contain residual anisotropies.

We start Section~\ref{sec:theory} by recalling the formalism of the MVs. We then introduce the notion of statistical variance in Riemannian spaces, from which the FVs are defined. In Section~\ref{subsec:toymodels} we analyze a mock Cold Spot model. In Section~\ref{sec:nulltests} we describe our pixel-based implementation of a null test of Gaussianity and isotropy, and in Section~\ref{sec:results} we present the results of our tests on 2018 Planck maps. We conclude in Section~\ref{sec:conclusions}.

\section{Multipole Vectors and Fréchet Vectors}\label{sec:theory}

The relationship between unit vectors and spherical harmonics appears to have many independent formulations in theoretical physics. The possibility of representing spherical harmonics by the direction of points on a sphere (or ``poles'', in Maxwell's jargon) was originally suggested by Gauss~\cite{hobson1931theory}. They were later implemented by Maxwell~\cite{maxwell1954treatise}, who referred to these directions as ``axes'', and used them in successive directional derivatives over the electric monopole to obtain an arbitrary charge multipole. In cosmology, multipole vectors were independently derived in~\cite{Copi:2003kt} (see also \cite{Katz:2004nj} for a polynomial approach), where they also received this name; they were later recognized as a re-derivation of Maxwell's result in~\cite{Weeks:2004cz}. However, the equivalence between the multipole moments of any harmonic function and unit vectors have long been recognized by relativists working with gravitational radiation, who use a symmetric and trace-free combination of $\ell$ vectors as a basis for symmetric trace-free tensors of rank $\ell$, which also form an irreducible representation of the rotation group $SO(3)$~\cite{sachs1961gravitational,pirani1964introduction,Thorne:1980ru,Blanchet:1985sp}. 
Fréchet vectors, on the other hand, are based on the concept of statistical moments of data points in curved spaces, a topic well known among statisticians~\cite{frechet1948elements}, but is essentially new in the context of CMB anisotropies.

In what follows, we quickly recap the equivalence between the harmonic and multipole vector representations of CMB. We then discuss the basics of statistical moments in Riemannian spaces, which serves as the foundation for our definition of the FVs.

\subsection{Multipole Vectors}

The multipole moments $\Delta T_\ell(\hat{\bs{n}})$ of the CMB temperature fluctuations, $\Delta T(\hat{\bs{n}})$, are commonly represented in terms of spherical harmonics as
\beq\label{eq:DT_ell_alms}
    \Delta T_\ell(\hat{\bs{n}}) = \sum_{m=-\ell}^{\ell} a_{\ell m} Y_{\ell m}(\hat{\bs{n}})\,.
\eeq
In the MV representation, the same object can be decomposed using a real scalar $\lambda_\ell$ and $\ell$ unit vectors $\boldsymbol{v}_i$ ($i=1,\cdots,\ell$) as 
\beq\label{eq:DT_ell_mvs}
    \Delta T_\ell(\hat{\bs{n}}) = \lambda_\ell\,(\boldsymbol{v}_1\cdot\boldsymbol{\nabla})(\boldsymbol{v}_2\cdot\boldsymbol{\nabla})\cdots(\boldsymbol{v}_\ell\cdot\boldsymbol{\nabla})\left.\frac{1}{r}\right|_{r=1}
\eeq
where $r=\sqrt{x^2+y^2+z^2}$. For a real field, $a^*_{\ell m}=(-1)^m a_{\ell, -m}$, which means that~\eqref{eq:DT_ell_alms} is described by $2\ell+1$ real numbers. Since~\eqref{eq:DT_ell_mvs} also contains $2\ell+1$ real numbers, one in 
$\lambda_\ell$ and $2\ell$ in $\{\bs{v}_1,\dots,\bs{v}_\ell\}$, these two representations are equivalent and an invertible transformation between them can be given in terms of symmetric and trace-free tensors (see Appendix~\ref{app:mathematics} for details). This transformation was explored in~\cite{Copi:2003kt} to build an algorithm from which the vectors $\boldsymbol{v}_i$ can be numerically obtained from the coefficients $a_{\ell m}$. However, this algorithm is inefficient for high $\ell$. An elegant and much more efficient algorithm was introduced in~\cite{Dennis_2004,Dennis_2005,Helling_2006}, where the MVs are given by the roots of a random polynomial having the $a_{\ell m}$s as coefficients:
\beq
\sum_{m=-\ell}^{\ell} \sqrt{\binom{2\ell}{\ell+m}} a_{\ell m}z^{\ell+m}\,.
\eeq
Being a polynomial of order $2\ell$, it has $2\ell$ complex roots. The coordinates of the MVs can be obtained from a given root through the stereographic projection $z_i=\cot(\theta_i/2)e^{i\phi_i}$. In the case of CMB, due to the reality condition of the $a_{\ell m}$s, it follows that these roots come in pairs $(z_i,-1/z_i^*)$. This implies that the MVs have an inherent reflection symmetry $(\theta_i,\phi_i) \leftrightarrow (\pi - \theta_i,\pi + \phi_i)$. Note that this symmetry could have already been anticipated from \eqref{eq:DT_ell_mvs}, which shows that not only the signs of the MVs are degenerate with that of the constant $\lambda_\ell$, but also their lengths is degenerate with the amplitude of this constant. For this reason, MVs are not usual vectors, but axes.\footnote{Technically, MVs are elements of the projective space $\mathbb{R}P^2$.} They have also been called \textit{headless vectors}, and in the following we may informally refer to them either as vectors or axes. The above algorithm also demonstrates why the MVs do not depend on the isotropic angular spectrum, since polynomials with coefficients $a_{\ell m}$ or $a_{\ell m}/\sqrt{C_\ell}$ have the same roots. This result has an important practical consequence: adding a Gaussian \textit{isotropic} noise with spectrum $N_\ell$ to the map does not change the MVs, since it boils down to shifting the angular spectrum to $C_\ell + N_\ell$. An implementation of this algorithm in Python, known as \texttt{polyMV}, was introduced in~\cite{Oliveira:2018sef}, and allows for a quick and reliable derivation of the MVs up to multipoles $\ell\sim10^3$.\footnote{\url{https://github.com/oliveirara/polyMV}.}

\subsection{Fréchet Vectors}

Given an input CMB map with resolution $\ell_\text{max}$, the total number of MVs grows as $\ell_\text{max}^2$. For maps with the angular resolution achieved by Planck and future CMB experiments, the number of vectors easily exceeds $10^6$. Thus, to test the symmetries of the CMB, it is important to look for ways to compress the physical information contained in these vectors. In other words, we need to devise a summary statistic of the MVs.

One possibility is to look for some definition of an average MV. However, the notion of vector addition is complicated by the lack of a global vector structure of the sphere, not to mention that the naive addition of MVs will not result in a single-valued operation, due to the freedom in choosing their global orientation. Concerning this last point, it is important to realize that any construction of summary vectors which combines the MVs but ignore their antipodes has the potential to introduce artificial biases in the final analysis~\cite{Abramo:2006gw}. Of course, the reflection symmetry of the MVs \textit{can} be explored for algebraic/numerical simplifications, just as the reality condition allows one to restrict the analysis to $a_{\ell m}$s for $m\geq 0$. We shall use this symmetry in our statistical test of Section~\ref{sec:nulltests}. Other available notions of summary statistics, such as the dot or cross-product of MVs, will generally result in a real number per multipole, more than one vector per multipole, or in vectors with no reflection symmetry.

But there is an elegant way out. Rather than regarding the MVs as lines passing through the sphere, we can regard them as pairs of antipodal points on the surface of the sphere. As it turns out, statistical notions of variance and mean in Riemannian spaces are well defined, and well known among statisticians. Given a set of $N$ random points on a metric space, the \textit{Fréchet variance} is defined as~\cite{AIHP_1948__10_4_215_0,dubey2019frechetanalysisvariancerandom}
\beq
\Psi(\bs{u}) \equiv \sum_{i=1}^{N}d^2(\bs{u},\bs{v}_{i})\,,
\eeq
where $d$ is the geodesic distance between the point $\bs{u}$ and the random points $\bs{v}_{i}$. This is a straightforward generalization of the notion of variance for data points on the real line, where the distance from the mean, $\Delta x_i = x_i - \bar{x}$, is replaced by the appropriate metric distance $d(\bs{u},\bs{v}_i)$. The vector $\bs{u}$, known as the \textit{Fréchet mean}, is the equivalent of $\bar{x}$, and is likewise defined as the point minimizing the (Fréchet) variance~\cite{buss2001spherical}.

The above definitions are completely general. In the case of the CMB, $d$ is the arc distance between two points on the unit sphere or $\gamma$. Since we are interested in capturing correlations among different CMB scales, it is interesting to introduce one variance per multipole:
\beq\label{eq:psi_ell}
\Psi_\ell(\bs{u}) \equiv \frac{1}{N}\sum_{i=1}^N\gamma^2(\bs{u},\bs{v}_{i})\,,\qquad
\cos\gamma = \bs{u}\cdot\bs{v}_{i}\,.
\eeq
This expression can be applied to any set of $N$ points on the sphere. However, when applied to the MVs, reflection symmetry must be respected, and in this case, we choose $N=2\ell$. Finally, we define the \textit{Fréchet vector} at each multipole, $\bs{u}_\ell$, as
\beq\label{eq:fvs-def}
\bs{u}_\ell \equiv \text{argmin}\,\Psi_\ell\,, \qquad \ell>1\,.
\eeq
From the reflection symmetry of the MVs, and using $\gamma(-\bs{a},\bs{b})=\gamma(\bs{a},-\bs{b})$, it follows immediately that $\Psi_\ell(\bs{u})=\Psi_\ell(-\bs{u})$, i.e., Fréchet vectors are also headless. This is an important property, as it allows us to use the very FVs in the mean \eqref{eq:psi_ell} and thus compute the mean of the mean. We will show an example in Section~\ref{subsec:toymodels}.

An important question that emerges refers to the uniqueness of the FVs.\footnote{We stress that by ``unique'' we mean uniqueness up to reflection symmetry.} A mathematically rigorous examination of this question is beyond our scope, and we point the reader to Refs.~\cite{buss2001spherical,eichfelder2019algorithm} for details. However, it is important to note that since $\Psi_\ell$ is continuous and the sphere is compact, it will always possess a minimum due to the extreme value theorem, although this minimum need not be unique. Indeed, more than one minimum is expected in a few and highly symmetric configurations. The first and obvious example occurs at $\ell=1$, in which case the MV and its antipode can be aligned with the $z$-axis. Since all points in the equator are equidistant to the north and south poles, FVs are degenerate in this case, which explains why we exclude $\ell=1$ in \eqref{eq:fvs-def}. At $\ell=2$, the two MVs will always form a great circle; since the north and south poles are the two antipodal points with minimum distance to this circle, there is only one FV in this case. A rare exception occurs when the two $\ell=2$ MVs are perfectly aligned, in which case the minimum is again degenerate. At $\ell=3$, the minimum is generally well defined, unless the three vectors are aligned or coincide exactly with the $x$-, $y$-, and $z$-axis. In the latter, we would have four FVs with coordinates $(\pm1/\sqrt{3},\pm1/\sqrt{3},\pm1/\sqrt{3})$. For $\ell=4,5,6$ there are no symmetric configurations of the MVs, and the FVs are unique if no alignments occur. Thus, quite generally, FVs will not be unique in situations where the MVs are arranged in exact symmetric configurations.  Luckily for us, we are interested in random MVs, so these cases are rare and can be easily flagged as numerical exceptions.

It is also instructive to study the limit in which the MVs continuously cover the sphere. The function $\Psi_\ell$ tends to a constant, and we can calculate this limiting value. For that, we partition the sphere into small squares of co-latitude $\Delta(\cos\theta)_i = 2/m$ and longitude $\Delta\phi_i = 2\pi/n$, where $m$ and $n$ are integers such that $mn=2\ell$. Since $\Psi_\ell$ is the same at any point in this limit, we choose $\bs{u}=\hat{\bs{z}}$, so that $\cos\gamma = v_{i}^z = \cos\theta_i$. Equation \eqref{eq:psi_ell} then becomes
\beq
    \Psi_\ell(\hat{\bs{z}}) = \frac{1}{4\pi}\sum_{i=0}^m\sum_{j=0}^n\theta_i^2\,\Delta(\cos\theta)_i\,\Delta\phi_j\,.
\eeq
In the limit $m,n\rightarrow\infty$, we finally find
\beq
    \Psi_\infty \equiv \frac{1}{4\pi}\int\theta^2\,{\rm d}^2\Omega = \frac{\pi^2-4}{2} \approx 2.9348.
\eeq

\begin{figure}
    \centering
    \includegraphics[width=\linewidth]{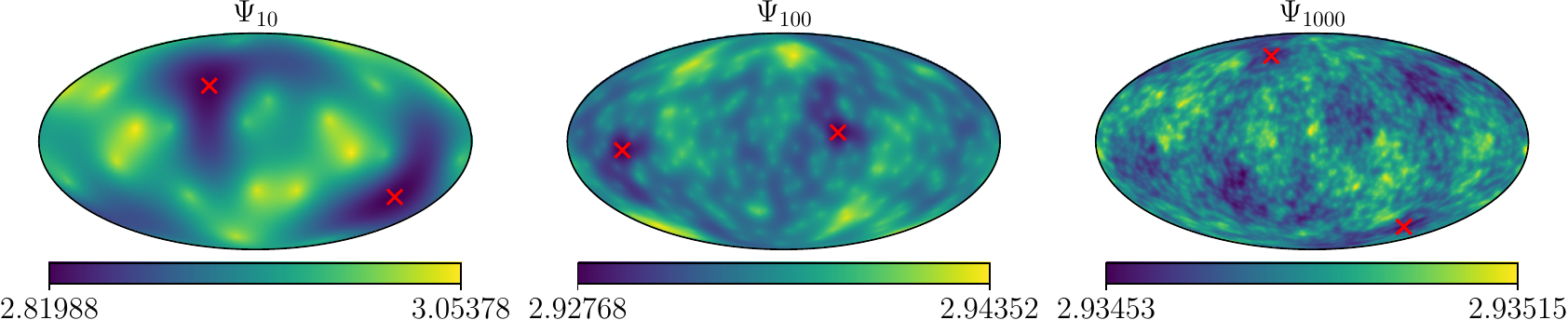}
    \includegraphics[width=\linewidth]{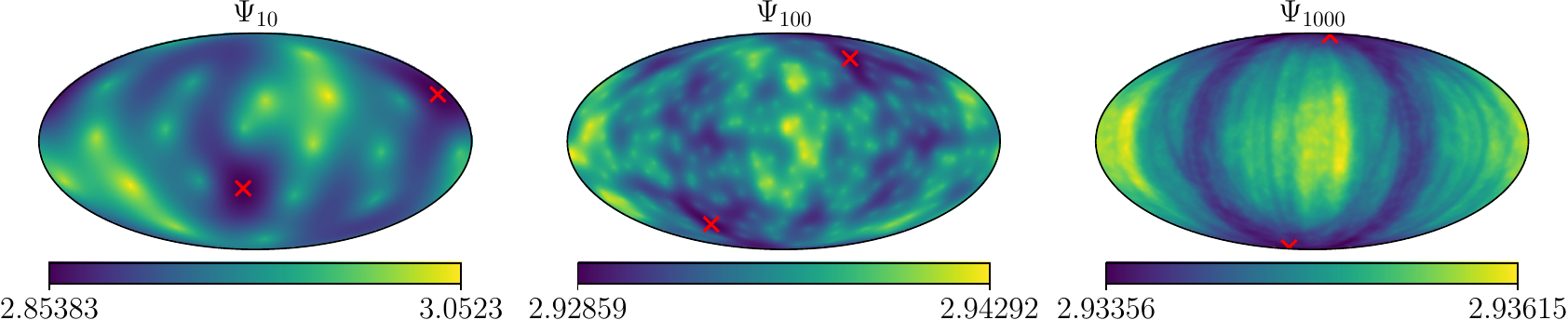}
    \caption{Fréchet variance of MVs at a few representative multipoles ($\ell_{\rm max} = 10, 100, 1000$). Their corresponding minima, represented by red crosses, define the FVs. Note how $\Psi_\ell$ approaches $\Psi_\infty \simeq 2.935$ for increasing values of $\ell$. 
    \textit{Top:} Gaussian and statistically isotropic (GSI) simulations.
    \textit{Bottom:} full-sky Planck 2018 \texttt{Commander} map.
    All plots in this paper are presented in galactic coordinates.
    }
    \label{fig:psi_maps_gsi}
\end{figure}

For random and uniformly distributed MVs in the range $\ell\in[2,2000]$, which is the range we consider in this work, the amplitude difference of two local minima of $\Psi_\ell$ are much larger than the numerical precision, and the global minimum can be easily found. This was confirmed both through a direct implementation of \eqref{eq:psi_ell}, as well as by using the \texttt{NLOPT}\footnote{\url{https://github.com/stevengj/nlopt}} optimization routine to find the local minimum of the function.   Figure \ref{fig:psi_maps_gsi} shows a few representative maps of the Fréchet variance, including the position of their minima, for both GSI simulations and \texttt{Commander} 2018 full-sky map, for $\ell_{\rm max} = 10$, 100 and 1000. Note that, as $\ell$ increases, the amplitude of the variances approaches the predicted value $\Psi_\infty$. The MVs of the 2018 \texttt{Commander} map used to produce these plots are displayed in Figure 1 of Ref.~\cite{Oliveira:2018sef}, and the striation pattern seen in $\Psi_{1000}$ was already observed there. It is known that the 2018 \texttt{Commander} map had a simpler treatment of the masked region, and the same pattern does not appear in the \texttt{Commander} 2015 map (see Figure 1 in \cite{Oliveira:2018sef}). We stress that these figures corresponds to full sky maps, which are known to be highly anisotropic. Here, this map was deliberately chosen to illustrate the marked anisotropy in the distribution of the variance maps.

\begin{figure}
    \centering
    \includegraphics[width=.94\linewidth]{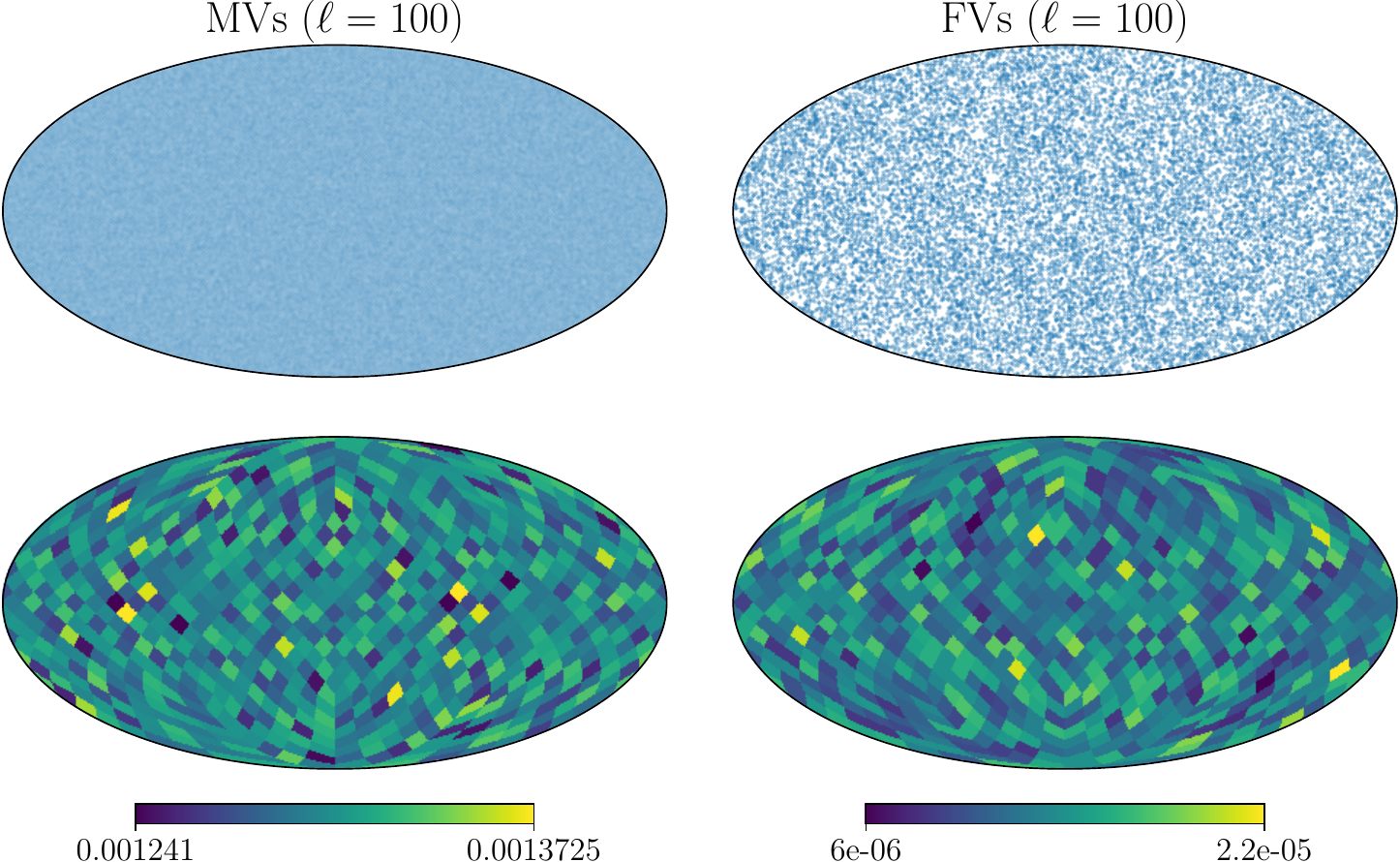}
    \caption{1-point distribution of the MVs (left) and FVs (right) for $\ell=100$, in real (upper line) and pixel spaces. The distribution in pixels is normalized, such that the frequencies summed over pixels gives 1. The vectors were drawn from $10^4$ simulations of Gaussian and statistically isotropic CMB maps using \texttt{polyMV}. For the lower panel we have adopted $N_{\text{side}}=8$.}
    \label{fig:1pcf}
\end{figure}

An important property of the FVs refers to their statistical distribution. Since for Gaussian and statistically isotropic CMB maps, MVs have a uniform 1-point distribution~\cite{bogomolny1992distribution}, we expect, by the definition of the FVs, that they are also uniformly distributed in this case. This is again confirmed by simulations, as shown in Figure \ref{fig:1pcf} for MVs and FVs extracted from GSI CMB maps at $\ell=100$, and shown both in real and pixel space. In the latter case, each pixel value refers to the (normalized) count of points (i.e., vectors) per pixel. The pixel-based representation of vector frequencies is a central part of our statistical analysis, and we shall return to this point later.

We note that due to the non-linear relation between the MVs and the $a_{\ell m}$s  (Eq.~\eqref{eq:alms2mvs}), vectors from the same multipole are in general non-Gaussian and correlated, even for $a_{\ell m}$s drawn from GSI realizations. In particular, all $\ell$-point cross-correlations of vectors from the same multipole are non-vanishing~\cite{Dennis:2007jk}. On the other hand, MVs corresponding to different $\ell$s are statistically independent as long as the $a_{\ell m}$ coefficients are independent. As for the FVs, since they correspond to one vector per multipole, they are also statistically independent for independent $a_{\ell m}$s. In this work, we will only study the 1-point distribution of the MVs and FVs to test the GSI hypotheses. The reader is warned that, for simplicity, we might occasionally speak of tests of isotropy (of the universe) or uniformity (of the vectors) when referring to the complete null test.

\section{Illustrating the power of FVs for a simulated Cold Spot}\label{subsec:toymodels}
To illustrate the power of the FVs, we will use them to reconstruct the axis of an anisotropic signal artificially introduced in isotropic simulations. As a toy model, we consider a simple cold spot, the angular profile of which can be obtained by fitting a function to the real cold spot (e.g., Figure 4 in~\cite{Owusu:2022etl}). For concreteness, we considered a model that mimics a Gaussian temperature profile with a 1$\sigma$ aperture of $\sim4^\circ$ and variable amplitude, which is then added to isotropic CMB realizations at galactic coordinates $(l,b)=(270^\circ,-15^\circ)$.\footnote{This position was chosen to ease visualization, but is otherwise arbitrary.} After adding this feature to an isotropic CMB map, MVs and FVs can be directly calculated using \texttt{polyMV}. The mathematical details of the implementation can be found in Appendix~\ref{app:cold_spot}.

\begin{figure}
    \centering
    \includegraphics[width=.95\linewidth]{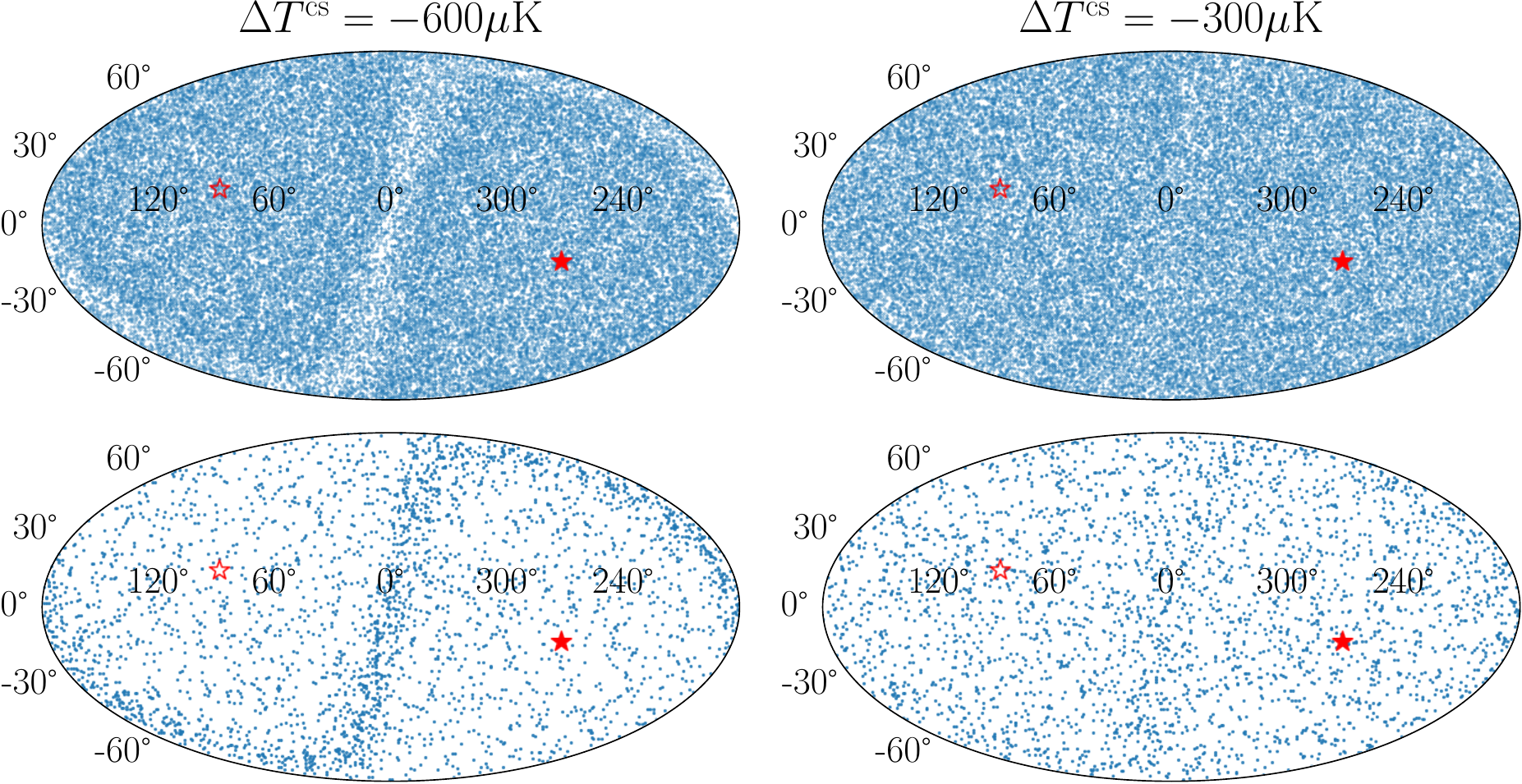}
    \caption{MVs (top) and FVs (bottom) for 50 realizations of CMB maps with a cold spot of $4^\circ$ 1$\sigma$ radius located at $(l, b)=(270^\circ, -15^\circ)$ (filled red stars). For convenience, we also show the antipode of the cold spot (empty stars). The cold spot models used have amplitudes of $-600\mu$K (left) and $-300\mu$K (right) --- see Appendix~\ref{app:cold_spot} for details. All vectors are in the range $\ell\in[2,30]$.}
    \label{fig:mvs-fvs-cspot}
\end{figure}

To estimate the accuracy of the FVs in reconstructing the axis of the cold spot, we generated 1000 GSI full-sky simulations to which the cold spot template was added. Since the anisotropies induced by the cold spot are restricted to the range $\ell\in[2,30]$ (Appendix~\ref{app:cold_spot}), we use MVs only from this interval. The MVs and FVs from 50 of these simulations are shown in Figure~\ref{fig:mvs-fvs-cspot} for two models with different cold spot amplitudes: $-600\mu $K and $-300\mu$K. Due to the non-linear relation between the $a_{\ell m}$s and the MVs, the effect of a cold-spot on the latter cannot be predicted analytically. However, the simulations displayed in Figure~\ref{fig:mvs-fvs-cspot} show that the MVs have a lower probability of pointing in the direction of an equatorial strip orthogonal to the cold spot axis, here represented by a filled red star. The FVs, instead, show a higher probability of pointing in at the cold spot region, since it corresponds to the positions that minimize the total distance to the MVs. This effect is more pronounced in the $-600\mu$K model and is barely visible in the $-300\mu$K case.\footnote{The effect is more easily seen by zooming out the page.} However, note that this is just a visual guide since the induced anisotropies are also hidden in the correlations of these vectors.

From a given cold spot simulation, we have 29 Fréchet vectors $\bs{u}_\ell$, with $\ell\in[2,30]$. Since they are headless, we can once again compute their mean by minimizing the Fréchet variance of these vectors. In other words, we can compute the FV of the FVs, which will result in a single vector per simulation; let us call these vectors $\bar{\bs{u}}_i$, with $i\in[1,1000]$. Given these vectors, we can again compute their Fréchet mean to estimate the \textit{global} mean, $\bar{\bar{\bs{u}}}$, as well as its dispersion. As a technical aside, we note that the global mean is computed from the $\bar{\bs{u}}_i$ by \textit{maximizing} Eq.~\eqref{eq:psi_ell}. This is only a convenience since a further minimization would return a vector that is orthogonal to $\bar{\bar{\bs{u}}}$. Finally, the one-sigma dispersion is computed as $\sqrt{\Psi_\ell(\bar{\bar{\bs{u}}})}$, which gives $33^\circ$ and $54^\circ$ for the models with amplitudes $-600\mu$K and $-300\mu$K, respectively.
\begin{figure}
    \centering
    \includegraphics[width=.95\linewidth]{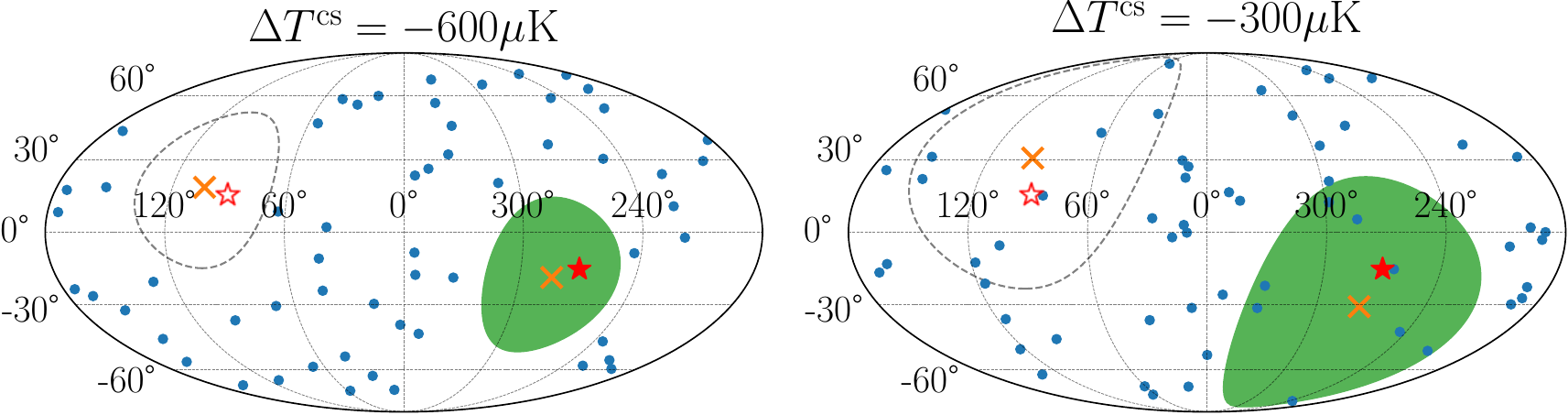}
    \caption{Reconstruction of a mock cold spot direction with the FVs for cold spot amplitudes $-600\mu$K (left) and $-300\mu$K (right). The blue dots represent the 29 FVs from one particular simulation $i$, whose mean is another headless vector $\bar{\bs{u}}_i$ (not shown). The global mean, obtained by computing the Fréchet mean over 1000 $\bar{\bs{u}}_i$s, is represented by an orange cross. We also show their 1$\sigma$ intervals (green) and corresponding antipodal intervals (dashed curves), which have radii of $33^\circ$ (left) and $54^\circ$ (right).}
    \label{fig:cs-reconstructed}
\end{figure}
The result of this analysis is shown in Figure~\ref{fig:cs-reconstructed} for the two models we considered, where we compare the reconstructed axis (orange cross) with the exact input axis (red star). As expected, the dispersion is smaller for the model with a stronger amplitude. 

We close this section with a few important remarks about this approach. First, note that since the FVs are headless by construction, we cannot reconstruct the \textit{direction} of the cold spot, but only its axis. On the other hand, since all the information of a CMB map is represented by the MVs set (i.e., the vectors plus the scalar $\lambda_\ell$), this direction \textit{can} be reconstructed, although we will leave this task for a future investigation, together with a proper assessment of the effect of masks and foregrounds in the determination of the cold spot. Second, the axis reconstruction we presented is, strictly speaking, not blind, since we have used \textit{a posteriori} information about the cold spot being zero for $\ell\gtrsim 30$. Indeed, had we included MVs from higher multipoles, we would be averaging over vectors uniformly distributed so that the reconstruction would have an increasingly larger variance. However, this does not invalidate our reconstruction, since, without this information, we could have run the test from $\ell=2$ up to some $\ell_{\text{max}}$ in several different bands of size $\Delta\ell$, until the variance of $\bar{\bar{\bs{u}}}$ attained a minimum. Indeed, this teaches us an important message when conducting isotropy tests with the MVs and FVs, which is to search for anisotropies across different multipole ranges. We now turn to an investigation of the symmetry properties of the CMB temperature fluctuations using these tools.

\section{Null test of Gaussianity and isotropy}\label{sec:nulltests}

The hypotheses of Gaussianity and isotropy are central to the standard cosmological model. Detecting possible deviations from this framework is of primary importance, as they could either indicate novel physical phenomena or reveal unaccounted systematics that may hinder our ability to identify such phenomena. We now turn to the investigation of the robustness of these hypotheses in the Planck data and under the lens of Multipole vectors and Fréchet vectors. 

As discussed in the previous section, MVs and FVs resulting from CMB temperature fluctuations should be uniformly distributed over the sphere if these hypotheses are correct. A straightforward way to test them is using a 1-point chi-squared test of statistical uniformity. This program was started in~\cite{Oliveira:2018sef}, where a coordinate-based approach was used to test the statistical uniformity of Planck vectors at all scales up to $\ell=1500$. Here, we extend this methodology by including several important enhancements in our statistical pipeline. First, we substitute a coordinate-based approach (in which the distribution of the components of the vectors, in a given coordinate system, is tested) for a pixel-based one (where we test the frequency of vectors per {\tt HEALPix}\footnote{\url{http://healpix.sourceforge.net}} pixel); this avoids the correlations between statistics in each coordinate and ensures our results are coordinate independent. This is also more appropriate for MVs and FVs, which are intrinsically (coordinate-independent) geometrical objects. We also improve our estimation of the covariance matrix entering the chi-square function, including a better assessment of potential biases entering its estimation and of its inverse. On the data side, we also include simulations of Planck's anisotropic instrumental noise in the estimate of covariance matrices. As is well known, Planck data is dominated by noise at multipoles $\ell\gtrsim1500$~\cite{Planck:2015mis}. With these realistic noise simulations, we also test the isotropy of the data up to $\ell=2000$.

The chi-squared test we shall perform has some important differences depending on whether we consider MVs and FVs, as we will explain below. Quite generally, though, it works as follows: given a CMB map at a fixed multipole $\ell$, and a {\tt HEALPix} map at resolution $\ns$, we count the frequency of vectors per pixel, $f_i$, with $i=1,\dots,\np$. Notice that since the frequency of vectors in pixels in the southern hemisphere is equal to the frequencies observed in the northern one, due to the reflection symmetry of the vectors, we can in practice restrict ourselves to the northern hemisphere, so that $\np=12\ns^2/2$, i.e., half of the standard {\tt HEALPix} value.\footnote{We stress that no bias is introduced here since we are just counting the independent frequencies. Other choices, such as left and right pixels, would give the same list of frequencies.} Using simulations, we then estimate the mean frequencies, $\bar{f}_i$, as well as their covariance matrix $C_{ij}$.\footnote{Since {\tt HEALPix} pixels are represented as one-dimensional arrays, $\bar{f}_i$ is computed as the arithmetic mean. Note also that $\bar{f}_i$ and $C_{ij}$ are computed separately for each $\ell$.} 
These quantities are then used to evaluate the reduced chi-square function of the input data, $f_i$, at the multipole $\ell$:
\beq\label{eq:chi2_ell}
    \chi^2_\ell(f_i) \equiv \frac{1}{\np}\sum_{i,j}^\np (f_i -\bar{f}_i)(C^{-1})_{ij}(f_j - \bar{f}_j)\,.
\eeq
$C^{-1}$ is the unbiased pseudo-inverse of the covariance matrix $C$, estimated from simulations following
\beq
    (C^{-1})_{ij} = \frac{N_\text{sims} - \np - 2}{N_\text{sims}-1}(C^{\rm P.I.})_{ij}\,,
\eeq
where $N_\text{sims}$ is the number of simulations. The multiplicative factor ensures that this estimator is unbiased~\cite{hartlap2007your}, and $C^{\rm P.I.}$ denotes the standard pseudo inverse of $C$. The use of a pseudo-inverse, rather than the usual inverse, is necessary since the covariance matrix is singular in our case. This happens because the frequencies are subject to the constraint $\sum_{i} f_i = \ell$, so that the frequency of any chosen pixel is a linear combination of the remaining $\np-1$ frequencies. In other words, any line of the covariance matrix is a linear combination of all the others, which proves that $C_{ij}$ is singular. For full-sky GSI maps, where isotropy is expected, we could in principle drop any chosen pixel from the frequency vector so as to obtain a non-singular covariance matrix of dimension $(\np-1)^2$. However, in real maps, some level of anisotropy is expected, either due to the introduction of a mask or due to residual foregrounds and instrumental noise, so this procedure could introduce some bias in the final results. Thus, in what follows, we stick to the definition~\eqref{eq:chi2_ell}. 

To estimate the different $f_i$ and $C_{ij}$ in Eq.~\eqref{eq:chi2_ell}, we have used 3000 simulations for each type of null hypothesis to which we want to compare the data (in our case, either unmasked full-sky maps, masked maps, or masked maps with anisotropic noise). An important issue at this point is the choice of an appropriate map resolution to compute $\bar{f}_i$ and $C_{ij}$. Small $\ns$ have pixels with large areas which could hide the correlation of vectors inside them. On the other hand, large $\ns$ is numerically expensive, since it requires a larger number of simulations to properly estimate these quantities. After some tests, we have found that a good compromise is achieved by choosing $\ns=8$ for vectors in $\ell\in[2,160]$ and $\ns=16$ for $\ell\in[161,2000]$. 

We remark that, for the maps to which anisotropic noise is added, the frequencies $\bar{f}_i$ will always present some degree of correlation with one another. This can result in reduced $\chi^2_\ell$ values which differ from unity, even when applied to simulations satisfying the null hypothesis. We dealt with this issue by generating an independent set of 2000 \textit{control simulations}, with the same features as the original 3000 simulations, which we used to estimate the distribution of~\eqref{eq:chi2_ell}. Using this numerical distribution, we compute the statistical significance of the data in comparison to the theoretical expectations.

\begin{figure}
    \centering
    \includegraphics[width=0.75\linewidth]{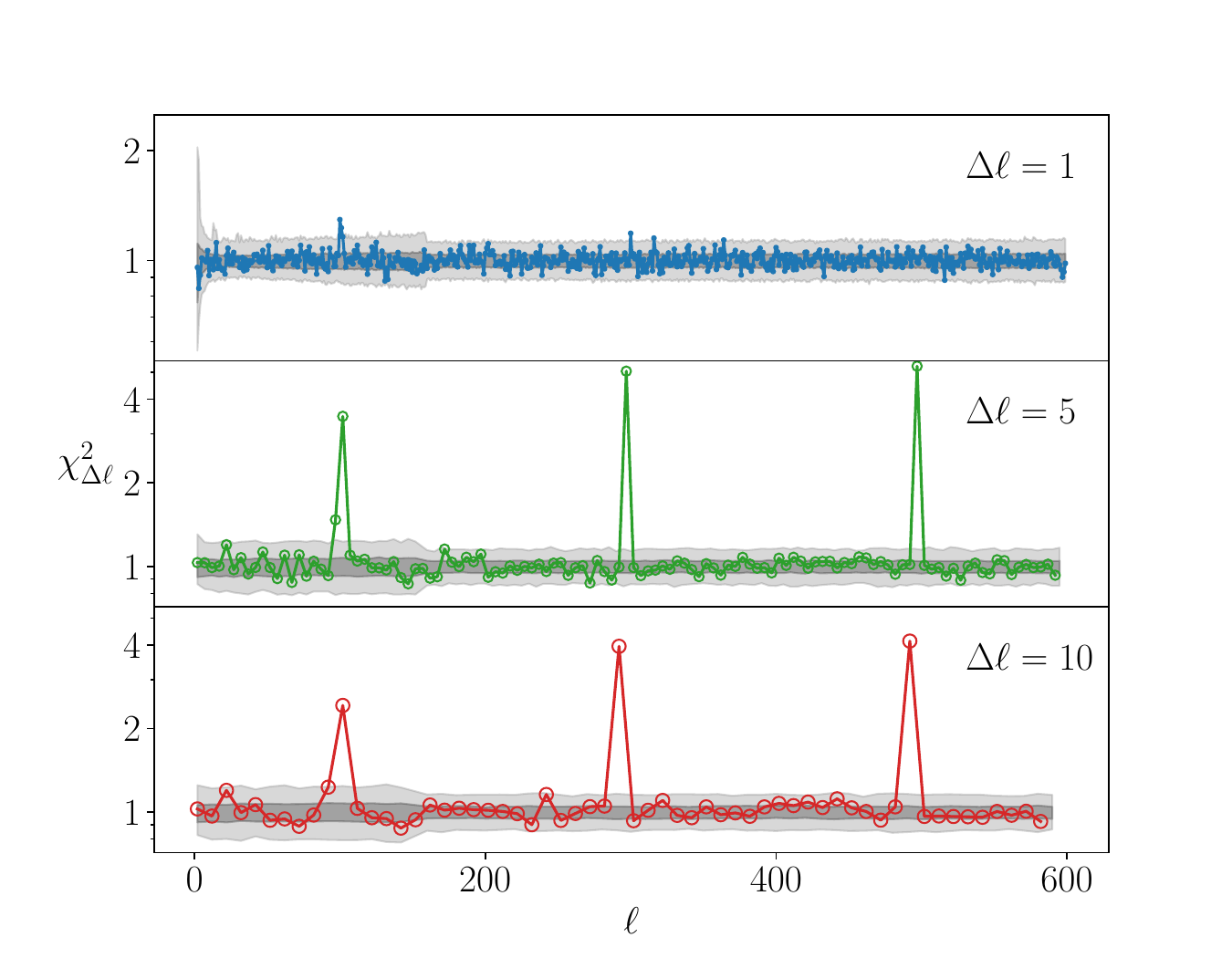}
    \caption{Chi-squared test of uniformity for mock maps where MVs are artificially repeated in $\delta\ell=5$ neighboring multipoles around the pivot scales $\ell_p=100$, 300, and 500. Each panel shows the result of a $\chi^2_{\Delta\ell}$ test using different $\Delta\ell$s, corresponding to joining MV sets in $\Delta\ell$ neighboring $\ell$s and testing for uniformity. Using $\Delta\ell=1$ the statistical fluke is unseen, while using $\Delta\ell=\delta\ell=5$  results in the highest significance, as expected. The dark and light gray bands correspond to regions of 1$\sigma$ and 3$\sigma$, respectively.}
    \label{fig:chi2_pseudomvs}
\end{figure}

As discussed in the end of Section~\ref{subsec:toymodels}, testing the uniformity of MVs in multipole bins of size $\Delta\ell$, rather than at each individual multipole, is particularly interesting, as it could help capture potential correlations between neighboring multipoles that would violate the GSI hypotheses. This requires a straightforward generalization of Eq.~\eqref{eq:chi2_ell} to a $\chi^2_{\Delta\ell}$ test, where vectors from an interval of multipoles (rather than from a single $\ell$) are used. We illustrate this by applying the $\chi^2_{\Delta\ell}$ test to a toy model, built as follows: starting with uncorrelated MVs extracted from a full-sky GSI simulation, we chose a pivot scale $\ell_p$ and make a copy of its $2\ell_p$ vectors. Next, for each of the scales $\ell_p-1$, $\ell_p-2$, ..., $\ell_p-\delta\ell$, where $\delta\ell$ is some arbitrarily chosen correlation range, we substitute the original vectors there by copies of the vectors from $\ell_p$, dropping in succession one single random MV (since we must have $2\ell$ MVs per $\ell$).  Individual MVs remain uniformly distributed at each $\ell$, but now are totally correlated across the range $\delta\ell$, and should appear in the $\chi^2_{\Delta\ell}$ when $\Delta\ell\approx\delta\ell$, signaling a break of statistical independence at those scales.  Figure~\ref{fig:chi2_pseudomvs} shows the results of this test where the input MVs have been copied at three pivot scales, $100$, $300$, and $500$, with a correlation range $\delta\ell=5$. The theoretical frequencies and covariances used in this test were derived from 1000 full-sky GSI simulations using a resolution $\ns$ as described above. Note how the chi-square test appears consistent with the null hypothesis when $\Delta\ell=1$ (falling within the $1.2\sigma$ interval), but is rejected for $\Delta\ell=5$ ($12\sigma$), as expected (see the Appendix~\ref{app:sigma-values} for how the confidence levels are computed). For $\Delta\ell=10$ we find a weaker rejection ($8\sigma$), since the signal is diluted by the inclusion of uncorrelated MVs. This shows that, in principle, the correct $\delta\ell$ used in this toy model could be reconstructed a posteriori as the one that maximizes the discrepancy. However, the main message is that testing the statistical distribution of the MVs with $\Delta\ell>1$ allow for a simple and straightforward test of anisotropies that affect neighboring multipoles.

For the FVs, the chi-squared test applies in the same way, except that Eq.~\eqref{eq:chi2_ell} is not computed at a single $\ell$. Since there is only one FV per multipole, cosmic variance is high, and a prohibitive number of simulations would be needed to estimate their mean frequencies and covariances in this case. Thus, we apply the test to all vectors from $\ell=2$ to a maximum multipole, which varies from $\ell_\text{max}=10$ up to $\ell_\text{max}=2000$. We then compute the significance for which the null hypothesis is excluded, as discussed in Appendix~\ref{app:sigma-values}.

\section{Results}\label{sec:results}

\begin{table}
\begin{centering}
\begin{tabular}{cc}
\toprule 
\textbf{Scales analyzed} & \textbf{Multipole range}\tabularnewline
\midrule 
Large & $2\leq\ell\leq31$\tabularnewline
Planck & $2\leq\ell\leq1500$\tabularnewline
All & $2\leq\ell\leq2000$\tabularnewline
\bottomrule
\end{tabular}
\par\end{centering}
\caption{CMB scales analyzed in this work.}\label{tab:scales}
\end{table}

Moving forward, we now present the results of our analysis using the 2018 Planck CMB maps. We shall focus primarily on temperature maps, which have a larger signal-to-noise ratio (S/N), and postpone the analysis of polarization to a future work. We consider all four Planck component-separation maps in the range $2\leq\ell\leq2000$: \texttt{Commander}, \texttt{NILC}, \texttt{SEVEM} and \texttt{SMICA}. Each of these maps is constructed using a distinct pipeline, but they all aim at removing as many foregrounds as possible. Modes in the range $2\leq\ell\leq1500$ were already investigated in~\cite{Oliveira:2018sef} using a simplified statistical and numerical pipeline. Although the noise power spectrum only surpasses the CMB one for $\ell > 1700$ for all four maps~\cite{Planck:2018yye}, the noise is already relevant at lower multipoles, especially for individual $a_{\ell m}$s. Indeed, in~\cite{Oliveira:2018sef}, a strong deviation of the GSI framework was detected at $\ell\gtrsim1300$, hinting at the presence of anisotropic noise and or residual foreground in the maps. Here we improve upon the analysis in~\cite{Oliveira:2018sef} by including the noise simulations. See also~\cite{Planck:2015mis}. which also allows us to probe the noise-dominated domain,  $1500<\ell\leq2000$. These comprise 300 simulations for each of the four component-separation maps. For this analysis, we worked with 2100 GSI simulations to estimate the vector frequencies and their covariance and repeated the use of each noise simulation 7 times. We also generate 1200 control maps similarly but repeating 4 times. We stress that each of these repeated noise simulations is added to independently generated GSI maps, so that the correlation induced by this repetition is small.

Finally, to mitigate possible selection biases, we quote our results in three well-motivated range of scales (Table~\ref{tab:scales}). These are divided into large scales ($2\leq\ell\leq31$),\footnote{The choice of 31 ensures there are a round 30 individual multipoles in this range. This allows more choices in terms of binning, which we will use below.} Planck scales ($2\leq\ell\leq1500$, where noise is still much smaller than the signal) and All scales $(2\leq\ell\leq2000)$. For the MV plots, we also show results in bins of 60 multipoles, where we either analyze each $\ell$ independently ($\Delta \ell = 1$) or in twelve groups of $\Delta \ell = 5$ or in four of $\Delta \ell = 15$. This separation in bins of 60 multipoles is used simply for visualization purposes, and in the global analysis we stick to $\Delta \ell$ of 1, 5, or 15 always. Regarding the underlying models to which the data are compared, we consider four separate cases: full-sky maps, masked maps, full-sky maps with noise, and masked maps with noise. Masked results, when quoted, were produced using the Planck Common Mask~\cite{Planck:2018yye}.

\subsection{Multipole vectors}
In cosmological investigations, CMB maps should always be used with a mask, so as to avoid foreground contamination from the galactic equator and other point sources. However, as a warm-up, it is interesting to start our tests with full-sky maps. Since they are known to contain residual foregrounds, anisotropies and non-gaussianities, they serve as validation of our method, which should flag the MVs from these maps as rejected by the null hypothesis (i.e., the hypothesis of uniformly distributed MVs --- see Figure~\ref{fig:1pcf}). 

\begin{figure}
    \centering
    \includegraphics[width=\linewidth]{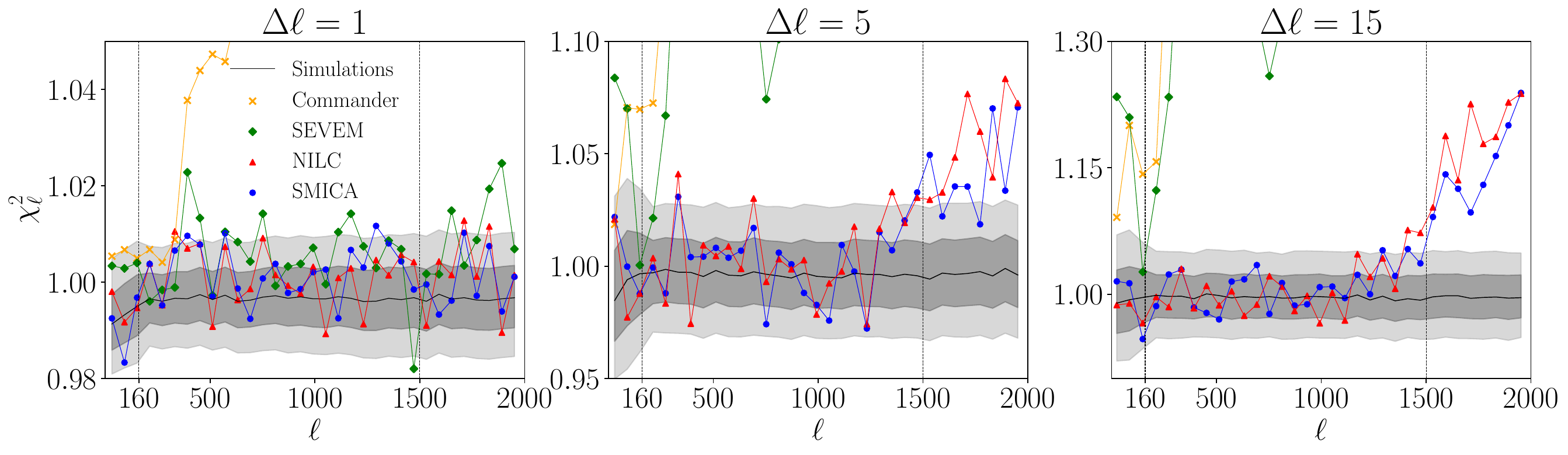}
    \caption{Reduced $\chi^2_\ell$ for full-sky Planck MVs compared against full-sky GSI simulations. The black line represents the mean values, while the grey bands the 1 and 2$\sigma$ regions derived from 2000 control simulations. The grey bands narrow down at $\ell=160$ due to the change from $\ns=8$ to $\ns=16$. 
    }
    \label{fig:chi2mvs_fullsky}
\end{figure}

\begin{table}
\begin{centering} 
    \begin{tabular}{ccccccc} 
    \toprule 
    {\textbf{Full-sky MVs}} & $\Delta\ell=1$ & $\Delta\ell=5$ & $\Delta\ell=15$ & $\Delta\ell=1$ & $\Delta\ell=5$ & $\Delta\ell=15$\\ 
    \midrule  Scales & \multicolumn{3}{c}{\texttt{Commander}} & \multicolumn{3}{c}{\texttt{NILC}} \\ 
    \midrule 
    {Large} & 0.8 & 1.8 & 1.8 & 0.4 & 1.4 & 0.8\\ 
    {Planck} & 52 & 261 & 291 & 1.0 & 1.6 & 1.2\\ 
    {All} &  76 & 348 & 383 & 0.7 & 6.1 & 10.1\\ 
    \midrule  & \multicolumn{3}{c}{\texttt{SEVEM}} & \multicolumn{3}{c}{\texttt{SMICA}}\\ 
    \midrule {Large} & 0.7 & 3.5 & 4.7 & 0.2 & 2.1 & 1.3\\ 
    {Planck} &  4.1 & 63 & 97 & 1.2 & 1.1 & 1.6\\ 
    {All} & 5.3 & 123 & 178 & 0.7 & 4.5 & 8.7\\ 
    \bottomrule 
 \end{tabular} 
\par\end{centering} 
\caption{$\sigma$-values of the data in Figure~\ref{fig:chi2mvs_fullsky} (more details in Appendix~\ref{app:sigma-values}). \texttt{Commander} and \texttt{SEVEM} are the most sensitive to the unmasked galactic regions, but when considering All scales and a correlation range of $\Delta\ell=5$ or 15, the GSI hypothesis is ruled out in all cases at high significance.}\label{tab:MVs-fullsky}
\end{table}

We show in Figure~\ref{fig:chi2mvs_fullsky} the results of the chi-squared test of uniformity applied to the full-sky Planck MVs. Note how the theoretical curve (black curve in Figure~\ref{fig:chi2mvs_fullsky}) is $\simeq1$ across all scales, which is consistent with the simulated frequencies being normally distributed for full-sky maps. As we can see, all Planck MVs present anisotropies in their distributions at varying scales. \texttt{Commander} and \texttt{SEVEM} are ruled out already at intermediate scales ($\ell\gtrsim160$). These two maps can be visually ruled out in the full-sky case (see Figure 1 in~\cite{Oliveira:2018sef}), so that our findings are expected. Regarding \texttt{SMICA} and \texttt{NILC}, their MVs seem surprisingly uniform for full-sky maps up to $\ell\simeq1500$, but are clearly rejected above these scales, due to the presence of anisotropic noise, residual foregrounds, or both.

The global goodness-of-fit of the data points in Figure~\ref{fig:chi2mvs_fullsky} with respect to the underlying theoretical model are shown in Table~\ref{tab:MVs-fullsky}, and for the three range of scales we are considering. It is interesting to note that \texttt{SMICA} and \texttt{SEVEM} 
appear as moderately ``anomalous'' at large scales ($\ell\in[2,31]$), being in tension with isotropy at 2.1 and $4.7\sigma$. These findings are more stringent than those found in~\cite{Oliveira:2018sef}, and seem to be on par with other claims of low-$\ell$ statistical anomalies~\cite{Schwarz:2015cma,Planck:2015igc,Muir:2018hjv,Jones:2023ncn}.

\begin{figure}
    \centering
    \includegraphics[width=\linewidth]{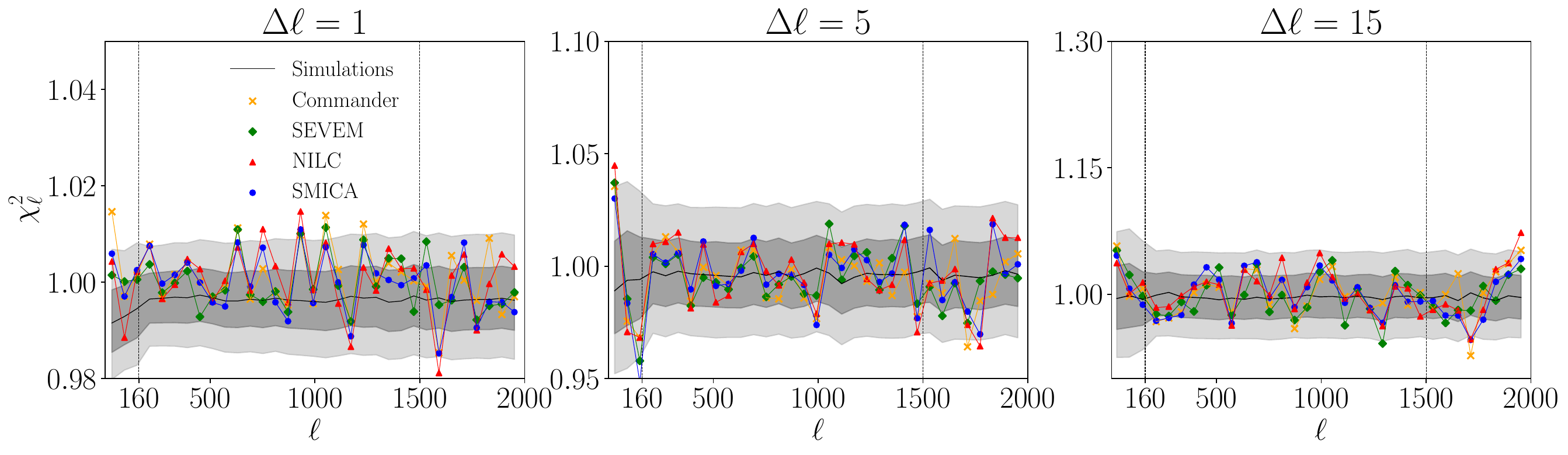}
    \caption{Same as Figure~\ref{fig:chi2mvs_fullsky}, but comparing masked CMB maps with masked simulations. The corresponding $\sigma$-values are shown in Table~\ref{tab:MVs-masked}.}
    \label{fig:chi2mvs_masked}
\end{figure}

\begin{table}
\begin{centering}
    \begin{tabular}{ccccccc} 
	\toprule {\textbf{Masked MVs}} & $\Delta\ell=1$ & $\Delta\ell=5$ & $\Delta\ell=15$ & $\Delta\ell=1$ & $\Delta\ell=5$ & $\Delta\ell=15$\\ 
    \midrule  Scales & \multicolumn{3}{c}{\texttt{Commander}} & \multicolumn{3}{c}{\texttt{NILC}} \\     
    \midrule {Large} & 1.8 & 1.6 & 1.8 & 1.0 & 2.3 & 1.0\\ 
    {Planck} &  2.2 & 0.1 & 0.8 & 1.6 & 0.3 & 1.1\\ 
    {All} & 1.2 & 0.03 & 1.0 & 0.8 & 0.2 & 1.2\\ 
    \midrule 
    & \multicolumn{3}{c}{\texttt{SEVEM}} & \multicolumn{3}{c}{\texttt{SMICA}}\\
    \midrule 
    {Large} &  0.7 & 1.9 & 1.6 & 1.4 & 1.6 & 1.5\\ 
    {Planck} & 1.0 & 0.2 & 0.5 & 1.3 & 0.2 & 0.9\\ 
    {All} &  0.4 & 0.04 & 0.4 & 0.3 & 0.1 & 0.6\\ 
    \bottomrule \end{tabular}
\par\end{centering}
\caption{$\sigma$-values of the data in Figure~\ref{fig:chi2mvs_masked}. All maps appear consistent with  isotropy in general, except \texttt{NILC} and \texttt{Commander} which show a $\gtrsim2\sigma$ tension at Large and Planck scales, respectively.}\label{tab:MVs-masked}
\end{table}

Next, we repeat the analysis for masked Planck maps: all maps entering the construction of Eq.~\eqref{eq:chi2_ell} are now masked before extracting the corresponding MVs. Our findings are shown in Figure~\ref{fig:chi2mvs_masked}, and their corresponding goodness-of-fit in Table~\ref{tab:MVs-masked}. 
As expected, the inclusion of the common mask renders all Planck maps consistent with Gaussianity and isotropy at Planck scales ($\ell\leq1500$). 
Surprisingly, Planck's  
MVs seem consistent with the GSI hypothesis even for $1500\leq\ell\leq2000$, where the S/N ratio is small. As we explain in Appendix~\ref{app:sens_frechet}, this simple 1-point function test of the MVs' distribution is not very sensitive to the presence of the anisotropic instrumental noise, which is more prominent at small scales. As we show below, this sensitivity is greatly improved using the Fréchet vectors.

\subsection{Fréchet Vectors}

\begin{figure}
    \centering
    \includegraphics[width=\linewidth]{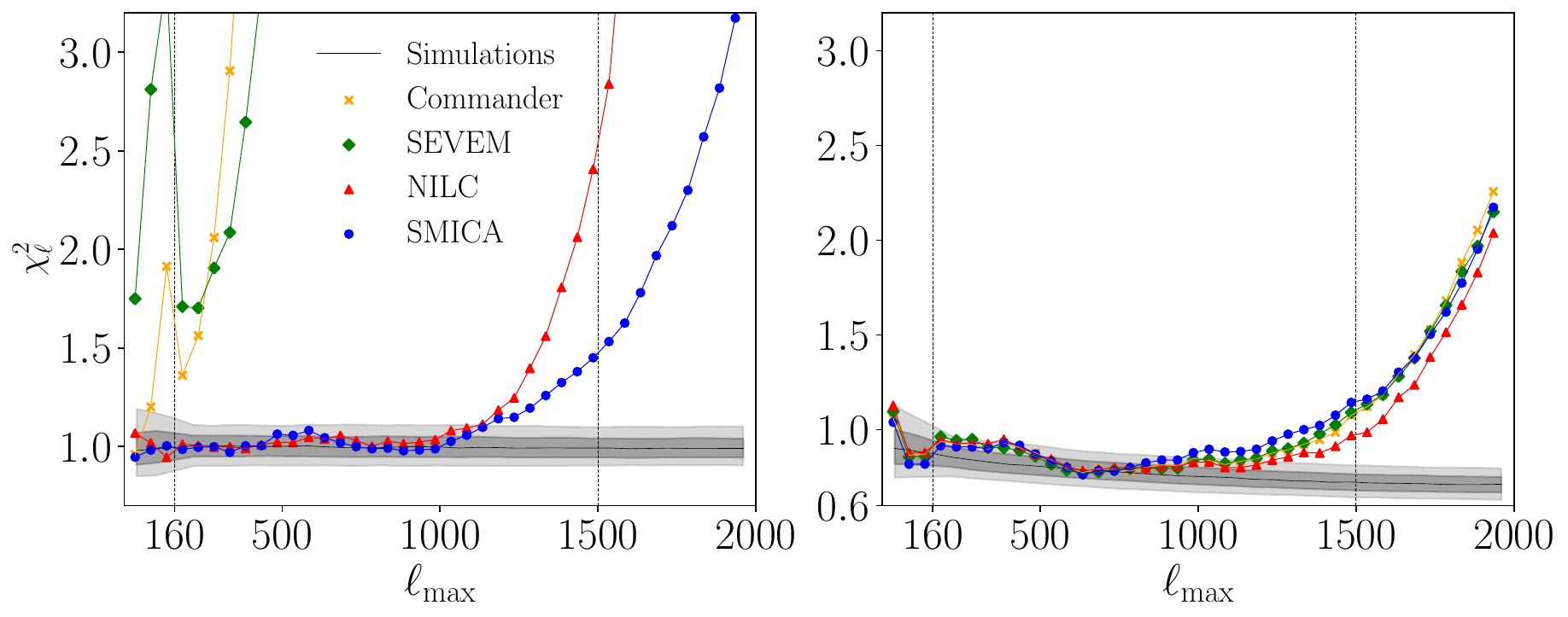}
    \caption{Reduced $\chi^2_\ell$ for Planck's FVs compared against full-sky (left) and masked (right) GSI simulations. Note that, for the FVs, $\chi^2_\ell$ is plotted against varying $\ell_{\text{max}}$. The gray bands correspond to 1 and $2\sigma$ variances derived from 2000 control simulations.}\label{fig:chi2fvs}
\end{figure}

Let us now analyze the behavior of Planck's FVs. As previously, we begin by comparing the vectors derived from full-sky Planck maps with those from full-sky GSI simulations and then proceed to compare masked versions of both data and simulations, assuming for now an isotropic noise. The outcomes of these analyses are illustrated in Figure~\ref{fig:chi2fvs} and Table~\ref{tab:chi2_fvs}. Recall that, as discussed in Section~\ref{sec:nulltests}, the chi-squared test for the FVs is applied to a range of multipoles $2\leq\ell\leq\ell_\text{max}$. As expected, all maps are ruled out in the full-sky case, with the null hypothesis being rejected with $8\sigma$ at $\ell\leq1500$, and with $27\sigma$ at $\ell\leq2000$ (both figures corresponding to \texttt{SMICA}). Note how \texttt{NILC} and \texttt{SMICA} maps, whose full-sky MVs appeared to satisfy the GSI hypothesis up to $\ell\leq1500$ (see Figure~\ref{fig:chi2mvs_fullsky} and Table~\ref{tab:MVs-fullsky}), are now ruled out in the same interval with more than $8\sigma$.


\begin{table}


\begin{centering}
    \begin{tabular}{ccccc} \toprule {\textbf{Full-sky FVs}} & \texttt{Commander} & \texttt{NILC} & \texttt{SEVEM} & \texttt{SMICA}\\ 
    \midrule {Large} & $0.6$ & $1.5$ & $5.1$ & $0.4$\\ 
    {Planck} &  $95$ & $19$ & $81$ & $8.0$\\ 
    {All} &  $102$ & $58$ & $ 86$ & $27$\\ 
    \midrule\midrule 
    {\textbf{Masked FVs}} & \texttt{Commander} & \texttt{NILC} & \texttt{SEVEM} & \texttt{SMICA}\\ 
    \midrule {Large} &  $1.1$ & $1.7$ & $1.8$ & $1.4$\\ 
    {Planck} &  $7.5$ & $5.3$ & $7.5$ & $8.2$\\ 
    {All} &  $21$ & $20$ & $21$ & $21$\\ 
    \bottomrule 
    \end{tabular}
\par\end{centering}
\caption{$\sigma$-values of the data in  Figure~\ref{fig:chi2fvs}.  All four component-separated and maps are rejected at Planck scales with more than $8.0\sigma$ (full-sky \texttt{SMICA}) and $5.3\sigma$ (masked \texttt{NILC}).}\label{tab:chi2_fvs}
\end{table}

The application of the common mask leads to substantially different results than those using MVs (Figure~\ref{fig:chi2mvs_masked} and Table~\ref{tab:MVs-masked}). At scales $\ell\leq1500$, we see that all component-separated masked maps are inconsistent with the null GSI hypotheses, the smallest and largest rejections corresponding to \texttt{NILC} ($5.3\sigma$) and \texttt{SMICA} ($8.2\sigma$), respectively. Clearly, extending the range to include the noise dominated regime $2\leq\ell\leq2000$, the rejection increases to around 20$\sigma$. In comparison, the test using MVs were not capable of detecting any anisotropy. As we explain in Appendix \ref{app:sens_frechet}, this is consequence of the MVs being binned in pixels, which erases most of their correlations at small scales. This corroborates our earlier claim that the test using FVs is more sensitive than the one with MVs to small angular variation.

\begin{figure}[t]
    \centering
    \includegraphics[width=\linewidth]{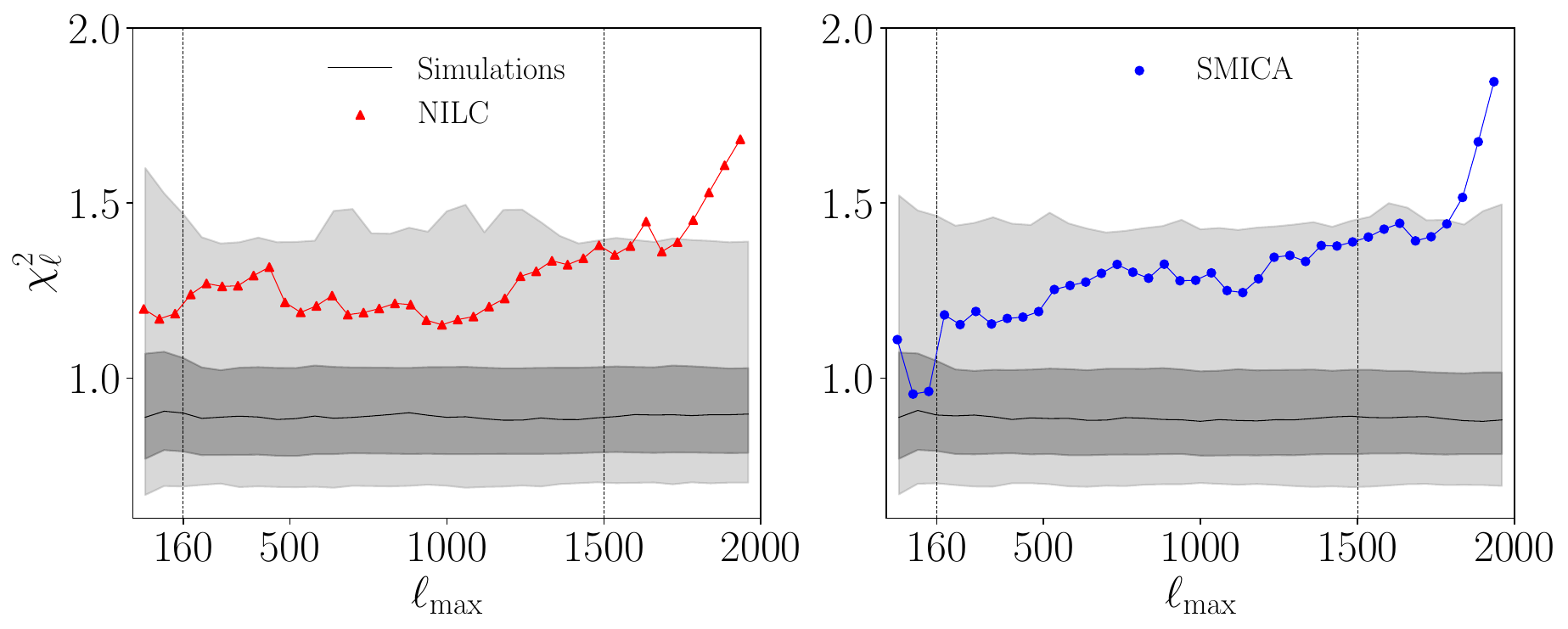}
    \caption{Reduced $\chi^2_\ell$ for Planck's FVs compared to GSI simulations with anisotropic noise and mask. The shaded regions correspond to $1\sigma$ and $2\sigma$ variances derived from control simulations.}
    \label{fig:chi2fvs_mask+noise}
\end{figure}

We now repeat the analysis by adding Planck's anisotropic noise simulations to our maps before they are masked. We use the 300 dx12\_v3 noise simulations provided by the Planck team for each of the four alternative CMB maps.\footnote{Available at the Planck Legacy Archive: \url{https://pla.esac.esa.int/}} From now on we will focus only on \texttt{NILC} and \texttt{SMICA} maps, since they are the least discrepant maps. The results are collected in Figure~\ref{fig:chi2fvs_mask+noise} and Table~\ref{tab:fvs_mask+noise}. As we can see, including the fully anisotropic noise simulations has a noticeable impact on the statistical significance of the FVs at scales $2\leq\ell\leq1500$, greatly improving their agreement with the GSI hypothesis. 
This confirms our earlier statements that the FVs are capable of blindly detecting the presence of anisotropies arising from noise even for $\ell\leq1500$, where the noise is still subdominant. We also note that there is no global hint of large-scale anomalies at $2\leq\ell\leq31$.

\begin{table}[t]
\begin{centering} 
\begin{tabular}{cccc} 
    \toprule {\textbf{Masked FVs + noise}} & \texttt{NILC} & \texttt{SMICA}\\ 
    \midrule {Large} &  $0.7$ & $0.8$\\ 
    {Planck} & $ 2.3$ & $2.1$\\ 
    {All} &  $3.5$ & $3.7$\\ 
    \bottomrule 
\end{tabular} \par\end{centering}
\caption{$\sigma$-values of the masked FVs in  Figure~\ref{fig:chi2fvs_mask+noise}.}
\label{tab:fvs_mask+noise}
\end{table}

There is, however, an important proviso in the above results. To minimize the variance in the analysis of the FVs, we computed the frequency of vectors per pixel and their covariances using all vectors in a range of multipoles from 2 to $\ell_{\text{max}}$. Since we treat the vectors in this range as equivalent, our results should be interpreted as reflecting an overall agreement between theory and data up to that maximum multipole. However, this approach dilutes possible anomalies which may affect more a specific range of multipoles. To address this, we repeated the analysis by binning the FVs into \textit{independent} intervals $\Delta\ell$ of sizes 25, 50, and 100. This choice of $\Delta\ell$s, which are multiples of each other, tries to avoid choosing specific ranges with too much arbitrariness. This binned analysis allows us to test the null hypothesis at different ranges of scales independently, while still avoiding too low statistics in each bin. Because the smallest $\Delta\ell$ we use imply in a single point at our Large scales case (see Table~\ref{tab:scales}), here we limit our analysis to the cases of Planck scales and All scales. 

\begin{figure}[t]
    \centering
    \includegraphics[width=\linewidth]{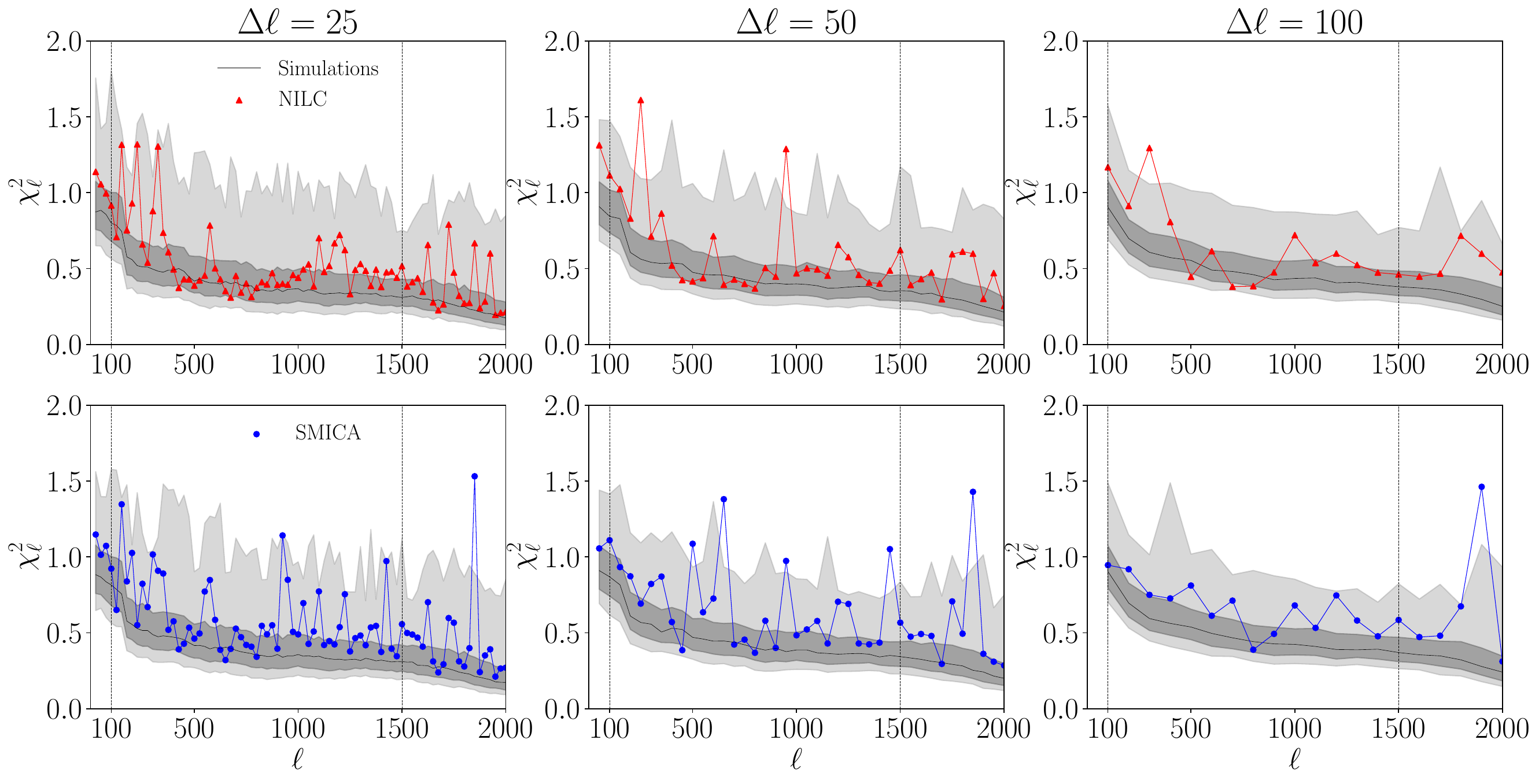}
    \caption{Same as Figure~\ref{fig:chi2fvs_mask+noise}, but now considering FVs binned in independent $\Delta\ell$ intervals of size 25, 50, and 100.}
    \label{fig:chi2fvs_mask+noise_binned}
\end{figure}


\begin{table}[t]
\begin{centering}
    \begin{tabular}{ccccccc}
    \toprule {\textbf{Masked FVs + noise}} & $\Delta\ell=25$ & $\Delta\ell=50$ & $\Delta\ell=100$ & $\Delta\ell=25$ & $\Delta\ell=50$ & $\Delta\ell=100$\\
    \midrule Scales & \multicolumn{3}{c}{\texttt{NILC}} & \multicolumn{3}{c}{\texttt{SMICA}} \\
    \midrule {Planck} &  $1.4$ & $1.9$ & $1.6$ & $2.7$ & $3.0$ & $1.8$\\
    {All} &  $1.5$ & $2.2$ & $2.1$ & $3.3$ & $3.7$ & $3.1$\\
    \bottomrule \end{tabular}
\par\end{centering}
\caption{$\sigma$-values of the masked FVs in Figure~\ref{fig:chi2fvs_mask+noise_binned}. \texttt{NILC} and \texttt{SMICA} are marginally consistent with GSI maps incorporating mask and anisotropic noise at scales $2\leq\ell\leq2000$.}
\label{tab:chi2fvs_mask+noise_binned}
\end{table}

The results, incorporating simulations with noise and mask, are shown in Figure~\ref{fig:chi2fvs_mask+noise_binned} and Table~\ref{tab:chi2fvs_mask+noise_binned}. Overall, the results are in reasonable agreement with those of Table~\ref{tab:fvs_mask+noise}. 
An inspection of Figures~\ref{fig:chi2fvs_mask+noise} and \ref{fig:chi2fvs_mask+noise_binned} suggests that these results are dominated by the signal in the range $1000\lesssim\ell\lesssim1500$. 
Of course, Planck data is known to have a residual noise contribution for  $1000\leq\ell\leq 1500$, and the above results may simply mean that the noise simulations may not be able to completely reproduce the anisotropies in the instrumental noise. Assuming that the dx12\_v3 noise simulations are reliable, these results point to the possible existence of residual foregrounds at arc-minute scales.


\section{Conclusions and perspectives}\label{sec:conclusions}

In this work, we made advances in the use of Multipole Vectors (MVs) to study anisotropies in the CMB. In particular, we showed that Fréchet Vectors (FVs), which are defined from the variance of MVs, allow for a more sensitive blind test of cosmic isotropy. As an illustration, we showed how FVs can be used to blindly detect the presence and position of a mock cold spot. Although the FVs themselves cannot reconstruct the direction of the cold spot due to their headless nature, all the information from a CMB map is retained in the multipole vector representation, suggesting that the direction can still be inferred. We will explore this issue in a forthcoming work.

We used both MVs and FVs to make 1-point statistical tests of isotropy using the Planck 2018 temperature maps (\texttt{Commander}, \texttt{NILC}, \texttt{SEVEM} and \texttt{SMICA}), improving upon the analysis of~\cite{Oliveira:2018sef} by including anisotropic noise simulations, more refined statistical tests, and by testing for anisotropies which affect simultaneously different scales. Overall, we found that the test using FVs is much more sensitive than the one with raw MVs, although we showed that the sensitivity of MVs can be improved by grouping multipole bins with $\Delta\ell>1$.

In particular, focusing on the scales probed well by Planck ($2\le \ell \le 1500$), the MV 1-point statistic found no clear anisotropy signal when Planck's common mask is applied to simulations and data. In fact, the MVs chi-square test performed well even at scales $\ell\geq1500$, where Planck maps have a low signal-to-noise ratio (Figure~\ref{fig:chi2mvs_masked} and Table~\ref{tab:MVs-masked}). As we demonstrate in the Appendix~\ref{app:sens_frechet}, this is not a limitation of the MVs, but follows from the pixel-based approach we adopt to bin the MVs into smaller datasets. Since the FVs chi-square test is not affected by this binning, they turn out more sensitive to the presence of anisotropic noise, and in this case we have included both the noise simulations and common temperature mask to our tests. Focusing on \texttt{NILC} and \texttt{SMICA}, which had a smaller amount of noise anisotropies, our results show small tensions with respect to the GSI hypotheses both at $\ell\leq1500$ ($\geq2.1\sigma$)  and $\ell>1500$ ($\geq3.5\sigma$) scales --- see Figure~\ref{fig:chi2fvs_mask+noise} and Table~\ref{tab:fvs_mask+noise}.

We performed a second analysis where we tested the distribution of the FVs in bins of $\Delta\ell$ equal to 25, 50 or 100, as shown in Figure~\ref{fig:chi2fvs_mask+noise_binned} and Table~\ref{tab:chi2fvs_mask+noise_binned}. We found similar results to those of Table~\ref{tab:fvs_mask+noise}. 

We also tested separately the large scales comprising $2\le\ell\le 32$, as since WMAP there have been many reports of anomalous statistical results in the CMB temperature maps. Our blind test using FVs found no global hint of anisotropies in this range of scales when analyzed together without further \textit{a posteriori} choices.

We stress that further analyses are needed to pinpoint the origin of the observed anisotropies in Planck's FVs, particularly at small scales where the noise is still subdominant. There are known mechanisms that could explain the discrepancies we found with respect to the Gaussian and isotropic model, such as Doppler aberration due to our peculiar motion relative to the CMB rest frame~\cite{Notari:2011sb}, or possible residual foregrounds, such as the recent claim of CMB contamination by nearby spiral galaxies not addressed by Planck’s common mask~\cite{Luparello:2022kqb}. Whether these mechanisms account for the anisotropies detected in this work is a topic for future investigation.

\acknowledgments{RGR is supported by CAPES (Coordenação de Aperfeiçoamento de Pessoal de Nível Superior). 
TSP is supported by FAPERJ (grant E26/204.633/2024), CNPq (grant 312869/2021-5) and Funda\c{c}\~ao Arauc\'aria (NAPI de Fen\^omenos Extremos do Universo, grant 347/2024 PD\&I). MQ is supported by the Brazilian research agencies FAPERJ (Fundação Carlos Chagas Filho de Amparo à Pesquisa do Estado do Rio de Janeiro, grant E26/201.237/2022), and CNPq (Conselho Nacional de Desenvolvimento Científico e Tecnológico). Numerical simulations were made using the computational resources of the joint CHE / Milliways cluster, supported by a FAPERJ grant E26/210.130/2023. The results of this work have been derived using {\tt GetDist}~\cite{Lewis:2019xzd}, {\tt HEALPix} ~\cite{Gorski_2005}, and {\tt Healpy} packages~\cite{Zonca2019}. All plots and figures were produced with \tt{Matplotlib}~\cite{Hunter:2007}}.

\appendix
\section{The formal relation between MVs and $a_{\ell m}s$}\label{app:mathematics}
Here we combine Maxwell's original derivation of the MVs with the language of spherical tensors to show the equivalence between Eqs.~\eqref{eq:DT_ell_alms} and \eqref{eq:DT_ell_mvs}. 

Consider two charges, at $\bs{r}=\bs{0}$ and $\bs{r}=\bs{s}_1$ and with potentials $\phi_0(\bs{r})=1/|\bs{r}|$ and $\phi_0(\bs{r}-\bs{s}_1)=1/|\bs{r}-\bs{s}_1|$. If $\phi_0(\bs{r}) - \phi_0(\bs{r}-\bs{s}_1)$ exists as $s_1\rightarrow 0$, we obtain the potential of a \textit{point dipole}, $\phi_1 = \lambda_1 \boldsymbol{v}_1\cdot\nabla(1/r)$, where the constant $\lambda_1$ accounts for the charges, and $\boldsymbol{v}_1$ is a unit vector parallel to $\bs{s}_1$. Now let a copy of the first point dipole, at 
$\bs{r}=\bs{s}_2$, approach the first at the origin. If 
$\phi_1(\bs{r})-\phi_1(\bs{r}-\bs{s}_2)$ exists as $s_2\rightarrow0$, we obtain the potential of a \textit{point quadrupole}, $\phi_2 = \lambda_2 (\boldsymbol{v}_1\cdot\nabla)(\boldsymbol{v}_2\cdot\nabla)(1/r)$. If an identical quadrupole approaches the first along $\bs{s}_3$, we obtain $\phi_3$, and so on. By induction, it follows that
\beq\label{eq:phi_ell}
\phi_\ell(\bs{r}) = \lambda_\ell (\boldsymbol{v}_1\cdot\nabla)\cdots(\boldsymbol{v}_\ell\cdot\nabla)\frac{1}{r}\,,
\eeq
Applying the gradients over $r = (x^ix_i)^{-1/2}$ and introducing unit vectors ${n_i=x^i/r}$, it follows that
\beq\label{eq:phi_ell_mvs}
\phi_\ell(\bs{r}) = (-1)^\ell\lambda_\ell(2\ell - 1)!!\,v_1^{\langle i_1}v_2^{i_2}\cdots v_\ell^{i_\ell \rangle }\,\frac{n_{i_1}n_{i_2}\cdots n_{i_\ell}}{r^{\ell+1}}\,,
\eeq
where repeated indices are summed, and $\langle\cdots\rangle$ represents a totally symmetric and trace-free combination of the enclosed indices.
This result applies to any harmonic function~\cite{courant2008methods}. For functions in the unit sphere, such as CMB temperature fluctuations, we set $r=1$ in~\eqref{eq:phi_ell_mvs}.
To establish the equivalence of \eqref{eq:phi_ell_mvs} with the usual harmonic expansion
\beq\label{eq:phi_ell_alms}
\phi_\ell(\bs{r}) = \sum_{m=-\ell}^\ell \frac{a_{\ell m}Y_{\ell m}(\hat{\bs{n}})}{r^{\ell+1}}, 
\eeq
we note that the spherical harmonics can be written in terms of $\ell$ unit vectors $n_i$ as
\beq\label{eq:ylm-n}
Y_{\ell m}(\hat{\bs{n}})={\cal Y}^{i_1 i_2\cdots i_\ell }_{\ell m}n_{i_1}n_{i_2}\cdots n_{i_\ell}\,,
\eeq
where the ${\cal Y}_{\ell m}^{i_1 i_2\cdots i_\ell}$ are totally symmetric and trace-free tensors
in the upper indices (known as spherical tensors) given by~\cite{Thorne:1980ru}
\begin{align}
{\cal Y}^{\ell m}_{i_{1}i_{2}\cdots i_{\ell}} & =C^{\ell m}\sum_{j=0}^{\lfloor(\ell-m)/2\rfloor}b^{\ell mj}\left(\delta^{1}_{(i_{1}}+i\delta^{2}_{(i_{1}}\right)\cdots\left(\delta^{1}_{i_{m}}+i\delta^{2}_{i_{m}}\right)\nonumber\\
 & \qquad\qquad\qquad\times\left(\delta^{3}_{i_{m+1}}\cdots\delta^{3}_{i_{\ell-2j}}\right)\times\left(\delta^{a_1}_{\ell-2j+1}\delta^{a_1}_{\ell-2j+2}\right)\times\dots\times\left(\delta^{a_j}_{i_{\ell-1}}\delta^{a_j}_{i_{\ell})}\right)\,.
\end{align}
Equating \eqref{eq:phi_ell_mvs} and \eqref{eq:phi_ell_alms} and using this result, it follows immediately that:\footnote{See \cite{Copi:2003kt} for an alternative derivation.}
\beq\label{eq:alms2mvs}
\lambda_{\ell}v^{\langle i_1}_1v^{i_2}_2\cdots v^{i_\ell\rangle}_\ell = \frac{(-1)^\ell}{(2\ell-1)!!}
\sum_{m=-\ell}^{\ell}a_{\ell m}{\cal Y}_{\ell m}^{i_1 i_2\cdots i_\ell}\,.
\eeq
As a quick check, let us consider the case $\ell=1$. Writing $Y_{1m}$ explicitly in terms of $n_i$ and comparing the result to \eqref{eq:ylm-n} reveals that ${\cal Y}^{i}_{10}=\sqrt{3/4\pi}\delta^i_3$, ${\cal Y}^i_{1-1}=\sqrt{3/8\pi}(\delta^i_1 - i\delta^i_2)=-({\cal Y}^{i}_{11})^*$. This gives, after absorbing a factor $\sqrt{4\pi/3}$ in $\lambda_1$:
\beq
\lambda_1 v^1_1 = \sqrt{2}a^\text{re}_{11}\,, \qquad 
\lambda_1 v^2_1= -\sqrt{2}a^\text{im}_{11}\,, \qquad 
\lambda_1 v^3_1 = -a_{10}\,.
\eeq
This agrees with Eq.(9) of Ref.~\cite{Copi:2003kt}, up to a global constant. 

The inverse relation can also be obtained by noting that the vectors $n_i$ obey an
orthogonality relation of the form
\beq\label{eq:ortho-n}
\int{\rm d}^{2}\Omega\,n_{i_{1}}n_{i_{2}}\cdots n_{i_{\ell}}=\begin{cases}
0 & \ell\;\text{odd}\\
\frac{4\pi}{\ell+1}\delta_{(i_{1}i_{2}}\cdots\delta_{i_{\ell-1}i_{\ell})} & \ell\;\text{even}
\end{cases}\,,
\eeq
where the parentheses denote a fully symmetric combination of indices. By equating \eqref{eq:phi_ell_mvs} and \eqref{eq:phi_ell_alms}, using the orthogonality of the spherical harmonics to isolate $a_{\ell m}$, followed by Eqs. \eqref{eq:ylm-n} and \eqref{eq:ortho-n}, we finally arrive at
\beq\label{eq:mvs2alms}
a_{\ell m} = 4\pi\frac{(-1)^\ell\ell!}{(2\ell+1)}\,\lambda_\ell v^{\langle i_1}_1v^{i_2}_2\cdots v^{i_\ell\rangle}_\ell {\cal Y}^{\ell m\,*}_{i_1 \cdots i_\ell}\,.
\eeq
Equations \eqref{eq:alms2mvs} and \eqref{eq:mvs2alms} establish the equivalence between the sets $\{\lambda_\ell,\bs{v}_\ell\}$ and $\{a_{\ell m}\}$. Since the objects ${\cal Y}^{\ell m}_{i_1\cdots i_\ell}$ form a basis for symmetric and trace-free tensors of rank $\ell$~\cite{Thorne:1980ru}, it follows from Eq.~\eqref{eq:alms2mvs} that the $a_{\ell m}$s are the coefficients of the tensor $\lambda_\ell v^{\langle i_1}\cdots v^{i_\ell\rangle}$ in this basis. Analogously, \eqref{eq:mvs2alms} shows that the tensor $a_{\ell m}$ has coefficients given by $\lambda_\ell v^{\langle i_1}\cdots v^{i_\ell\rangle}$ in the reciprocal basis. 

\section{Cold Spot}\label{app:cold_spot}

\begin{figure}\label{fig:csprofile}
\begin{center}
\includegraphics[scale=0.46]{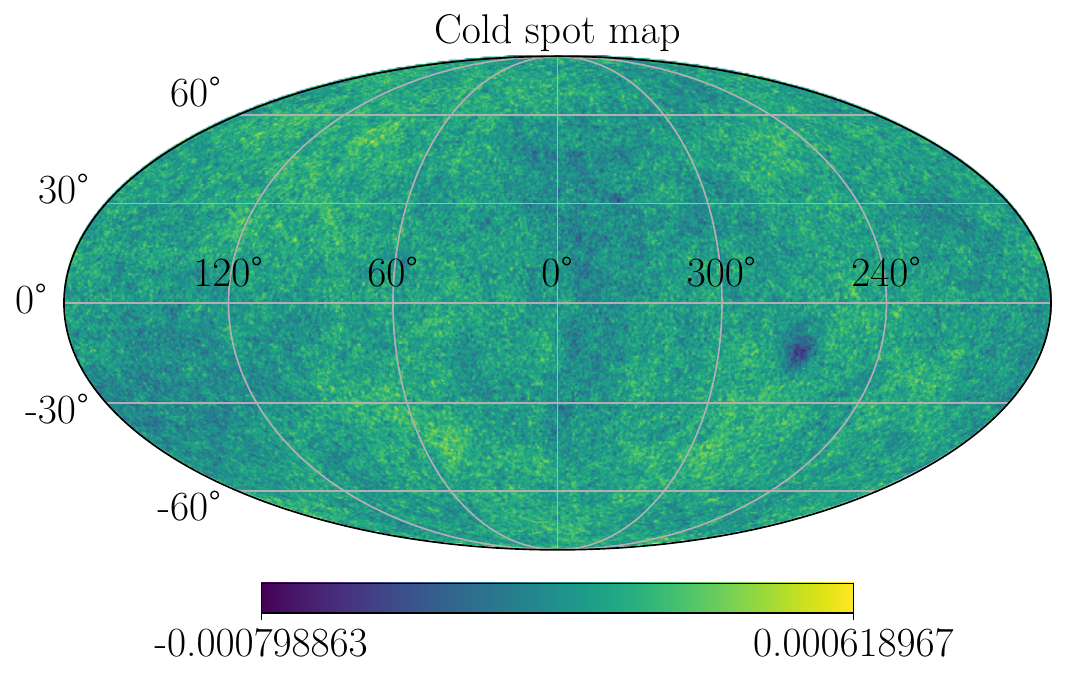}\;
\includegraphics[scale=0.46]{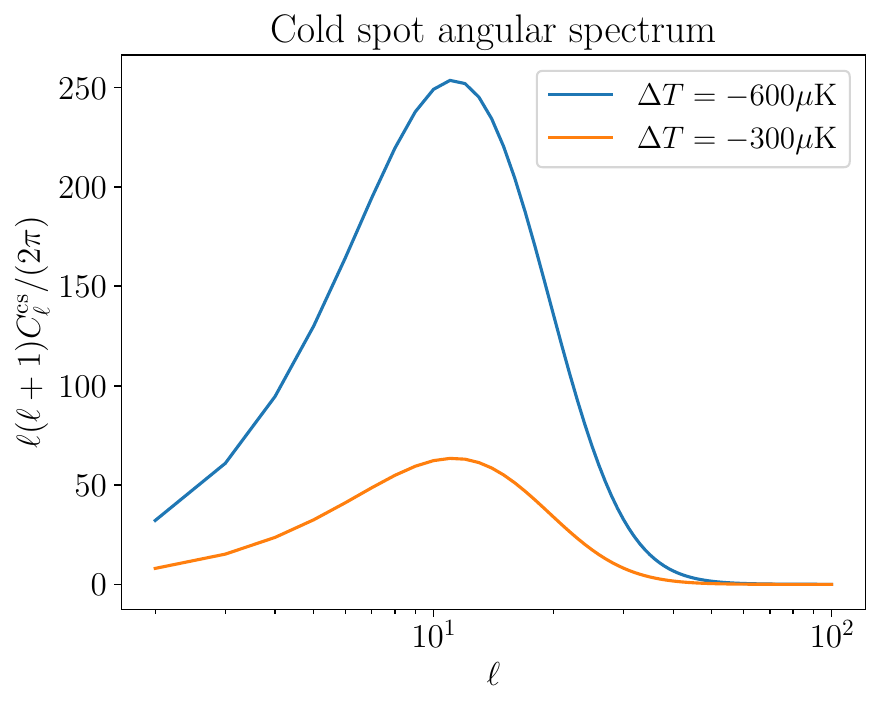}
\caption{Left: Gaussian and isotropic map with a cold spot modeled by eq.~\eqref{eq:csprofile} with $A=-600\mu$K and placed at
$(l,b)=(270^\circ,-15^\circ)$. Right: cold spot angular spectrum.}
\end{center}
\end{figure}

The simulations used in Section \ref{subsec:toymodels} were modeled with a phenomenological cold spot profile of the form
\begin{equation}\label{eq:csprofile}
\Delta T^{\mathrm{cs}} = A(-1 + \tanh(10\times\theta))^{1.5}\,,
\end{equation}
and placed at the north pole, where $A$ is a constant amplitude. This function mimics a Gaussian profile with a standard deviation $\sigma_\theta\sim 4^\circ$, which is slightly larger than the observed cold spot, with $\sigma_\theta\sim3.5^\circ$~\cite{Owusu:2022etl}. Maps of the cold spot were generated at resolution $\ns=512$, and then added to random GSI maps, generated from a seed $C_\ell$ (using Planck's best fit parameters), at the same resolution. The location $(l, b)=(270^\circ,-15^\circ)$ used in our simulations can be obtained using Healpy rotation tools directly on the maps. Figure~\ref{fig:csprofile} shows a sample map for $A=-600\mu$K, together with the power spectrum of the cold spot. Note that it peaks at $\ell\sim10$, and is negligible for $\ell\gtrsim30$.

\section{Computing $\sigma$-values}\label{app:sigma-values}

\begin{figure}
    \centering
    \includegraphics[width=0.5\linewidth]{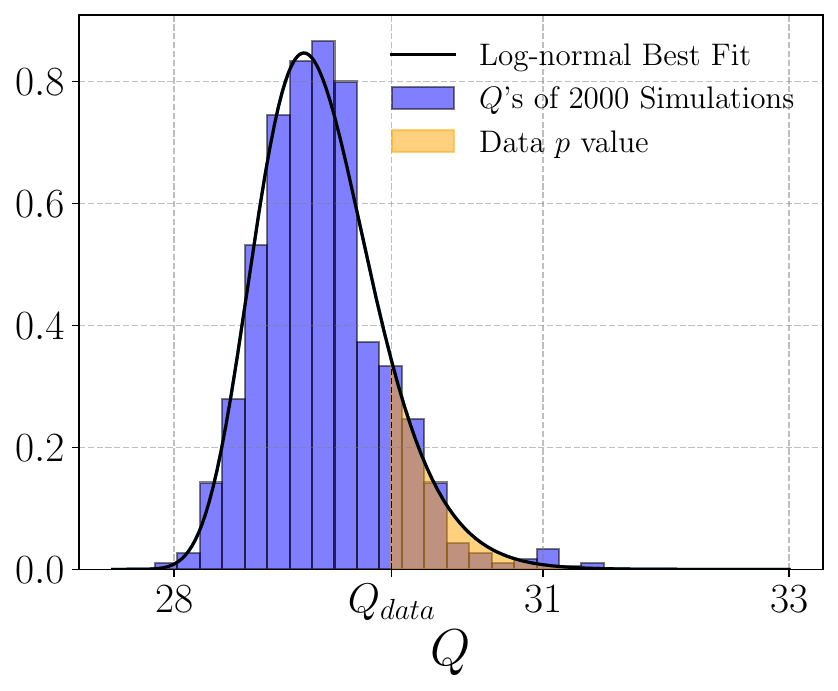}
    \caption{
    Histogram of $Q$'s calculated from 2000 simulations and the best-fitting continuous probability distribution function (PDF). In this example, we calculate the $p$-value of the \texttt{NILC} map, which is the area below the best fit PDF from the $Q_{\textit{data}}$ to $\infty$. For this plot, only Large scale multipoles were used (see Table~\ref{tab:scales}).}
    \label{fig:hist_pvalues}
\end{figure}

For a given range of multipoles, the algorithm described in Section~\ref{sec:nulltests} will generate a list of reduced chi-square values for each $\ell$ in this range. To transform this list into one number characterizing the global agreement between the data and the null hypothesis, we proceed as follows: from the list of $\chi^2_\ell$, we build the total chi-square, defined as
\beq\label{eq:Q}
    Q \equiv \sum_{\ell=2}^{\ell_\text{max}} \chi^2_\ell\,.
\eeq
$Q$ will also be $\chi^2$ distributed if the individual $\chi^2_\ell$ are, only with a different number of degrees of freedom. This is a good description for full-sky and masked GSI simulations, where the frequencies are (approximately) normally distributed. For more realistic maps, such as maps with mask and anisotropic noise, we use our 2000 control simulations to obtain a fit for the probability distribution of $Q$. We have checked that, for many situations, $Q$ is well approximated by a log-normal distribution. From $Q$'s probability distribution we get the $p$-value of each map, as shown in Figure~\ref{fig:hist_pvalues}. From the $p$-values we calculate the $\sigma$-values with
\beq\label{eq:sigma}
    \sigma{\text{-value}} = \sqrt{2} \erf^{-1}(1 - p{\text{-value}}).
\eeq
The description above works for all MVs analysis, and for the FVs analysis of Table~\ref{tab:chi2fvs_mask+noise_binned}. For the FVs analysis of Tables~\ref{tab:chi2_fvs} and \ref{tab:fvs_mask+noise}, since we use all the vectors from multipoles $\ell=2$ to $\ell_{\text{max}}$ for the calculation of a single chi-square, the total chi-square is obtained as: $Q = \chi^2 _{\ell_\text{max}}$.
Finally, for the $\chi^2$ plots, we estimate the distribution from our 2000 control simulations using the  Kernel Density Estimation technique available in the $\tt{GetDist}$ Python Package \cite{GetDist}. The $1\sigma$ and $2\sigma$ regions from the distribution correspond to the High-Density Intervals (HDI).

\section{On the sensitivity of the chi-square test using FVs}\label{app:sens_frechet}

Our tests using FVs result in overall more sensitivity compared to the same tests applied to the MVs. This difference is not intrinsic to the FVs -- after all, they contain less information than the MVs -- but rather results from a loss of statistical power in the MVs' chi-square test due to their binning in pixels. By replacing all MVs inside a given pixel by a single number (their frequency $f_i$ at that pixel), we lose much of  information in their correlations. FVs, on the other hand, are computed directly from the (unbinned) MVs, and the correlations are takien into account when computing this single vector, which is only then binned into a pixel frequency needed to perform the test.

\begin{figure}[t]
    \centering
    \includegraphics[width=\linewidth]{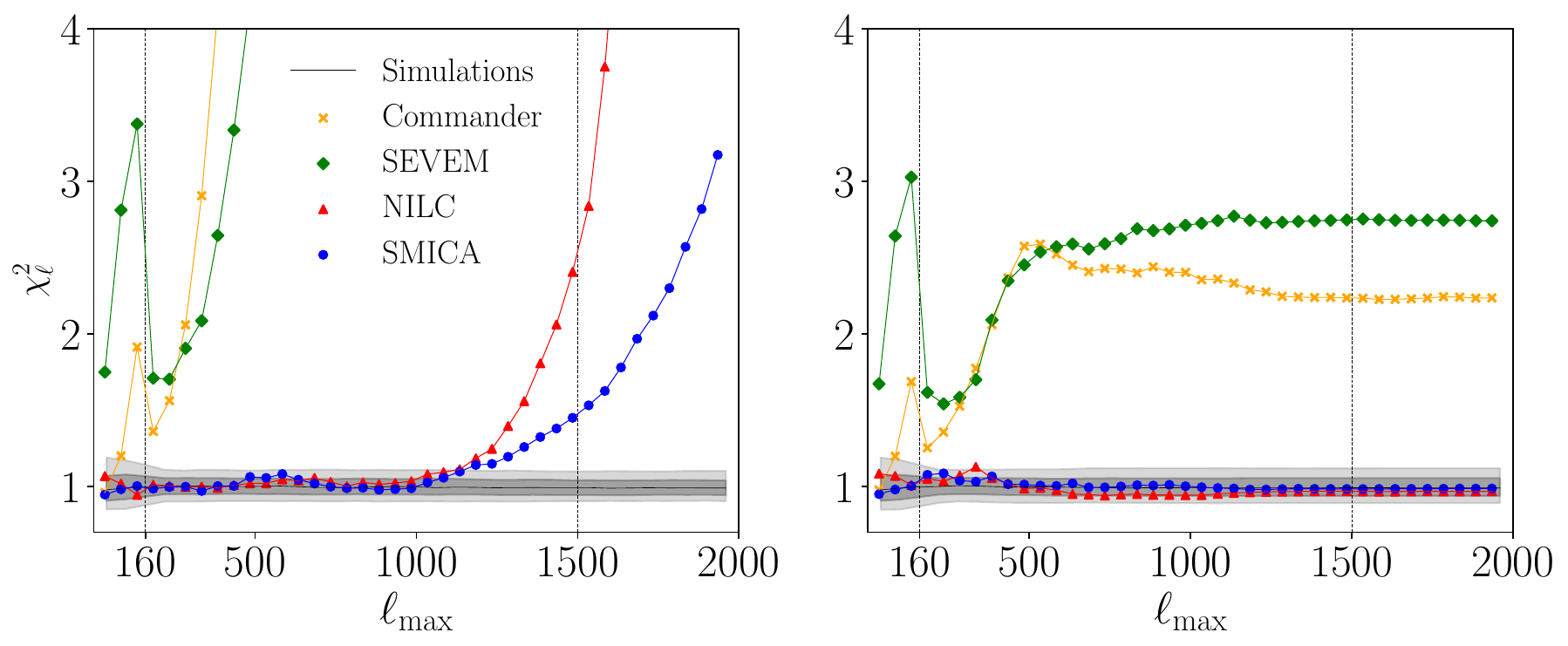}
    \caption{Comparison between the $\chi^2$ test using the original FVs, and a new set computed from binned MVs (see text). \emph{Left:} original FVs. \emph{Right:} ``binned'' FVs. Binning the MVs before computing the FVs translates into significant loss of statistical power.}
    \label{fig:fvs_sensitivity}
\end{figure}

\begin{table}
\begin{centering}
    \begin{tabular}{ccccc} \toprule {\textbf{Full-sky FVs}} & \texttt{Commander} & \texttt{NILC} & \texttt{SEVEM} & \texttt{SMICA}\\ 
    \midrule {Large} & $0.5$ & $1.4$ & $5.1$ & $0.4$\\ 
    {Planck} &  $14$ & $0.4$ & $18$ & $0.5$\\ 
    {All} &  $14$ & $0.4$ & $18$ & $0.6$\\ 
    \bottomrule 
    \end{tabular}
    \par\end{centering}
    \caption{$\sigma$-values of the data shown in the right panel of Figure~\ref{fig:fvs_sensitivity}.  These should be compared to the $\sigma$-values given in Table~\ref{tab:chi2_fvs}, corresponding to data in the left panel of Figure~\ref{fig:fvs_sensitivity}.
    \label{tab:chi2_fvs_binned}}
\end{table}

As an illustration of the sensitivity loss due to the binning, we have repeated the full-sky FVs chi-square test using FVs extracted not from the original full-sky MVs, but from a binned version of them. To construct this binned set, we first ask how many MVs of the original set fall in a given pixel $i$, for a given $\ns$. We then generate $f_i$ copies of an arbitrarily chosen vector of this subset, and repeat this process for all pixels.\footnote{Note that the binning method here bears no connection with the binning leading to Figure~\ref{fig:chi2_pseudomvs}.} This procedure erases the correlations of MVs inside pixels, and should have an impact in the new FVs. As a consequence, the resulting chi-square test should be strongly affected, losing most of their statistical power. In Figure \ref{fig:fvs_sensitivity} below we compare the original test (left) with the same test applied to new FVs (right), and in Table~\ref{tab:chi2_fvs_binned} we show the new $\sigma$-values. As we can see, the $\sigma$-values of all Planck maps are significantly affected in all range of scales compared to those in Table~\ref{tab:chi2_fvs}. NILC and SMICA maps are now consistent with the GSI hypothesis at $\lesssim 2\sigma$ at all scales, while SEVEM and Commander, although still discrepant, are now ruled out with less confidence (recall that these are all full-sky maps). Note also that these new values are now comparable to those of the full-sky MVs for $\Delta\ell=1$, shown in Table~\ref{tab:MVs-fullsky}.

\bibliographystyle{JHEP}
\phantomsection\addcontentsline{toc}{section}{\refname}\bibliography{references-fvs-paper}

\end{document}